\documentclass[aps,twocolumn,amsmath,amssymb,pra,groupedaddress]{revtex4-2}

\usepackage{xcolor}
\usepackage{bbm}
\usepackage{hyperref}
\usepackage{graphicx}
\usepackage{subcaption}
\usepackage{adjustbox}
\usepackage[normalem]{ulem}
\usepackage{wrapfig}
\usepackage{ulem}

\begin{document}

\title{Variational Tensor Network Simulation of Gaussian Boson Sampling and Beyond}

\author{Jonas Vinther}
\affiliation{Department of Computer Science, University of Copenhagen, DK}
\affiliation{Niels Bohr Institute, University of Copenhagen, DK}
\email[Contact author: ]{jonas.vinther@nbi.ku.dk}

\author{Michael J. Kastoryano}
\affiliation{Department of Computer Science, University of Copenhagen, DK}

\date{\today}

\begin{abstract}
The continuous variable quantum computing platform constitutes a promising candidate for realizing quantum advantage, as exemplified in Gaussian Boson Sampling. While noise in the experiments makes the computation attainable for classical simulations, it has been suggested that the addition of non-linear elements to the experiment will help retain the quantum advantage. We propose a classical simulation tool for general continuous variable sampling problems, including Gaussian Boson Sampling and beyond. We reformulate the sampling problem as that of finding the ground state of a simple few-body Hamiltonian. This allows us to employ powerful variational methods based on tensor networks and to read off the simulation error directly from the expectation value of the Hamiltonian.
We validate our method by simulating Gaussian Boson Sampling, where we achieve results comparable to the state of the art. We also consider a non-Gaussian sampling problem, for which we develop novel local basis optimization techniques based on a non-linear parameterization of the implicit basis, resulting in high effective cutoffs with diminished computational overhead.
\end{abstract}
 
\maketitle

\section{Introduction} 

Continuous variable (CV) quantum systems constitute a conceptually different framework for quantum computation as compared to discrete, qudit based architectures \cite{RevModPhys.77.513}\cite{PhysRevA.79.062318}\cite{Bourassa2021blueprintscalable}. An essential milestone that has been pursued over the last decade, is the proof of a quantum advantage \citep{v009a004}. For CV systems, one of the candidates for such an experiment is the Gaussian Boson Sampling (GBS) problem \citep{PhysRevLett.119.170501}\cite{doi:10.1126/sciadv.abi7894}.
Standard GBS is based on squeezed vacuum states of light that passes through rotation gates and beam-splitters and are subject to either threshold or Photon Number Resolved (PNR) detection at the end of the circuit \cite{PhysRevA.98.062322}\cite{ZHONG2019511}.
Calculating a probability for obtaining one of the many combinations of photon number measurements is an exponentially hard computational problem in the number of excitations \cite{doi:10.1126/sciadv.abl9236}\cite{PhysRevResearch.2.023005}\cite{Gupt2020}.

Previous attempts at handling this curse of dimensionality in simulating boson sampling, includes the approximate algorithm of Ref. \cite{Neville2017} which is based on Metropolis independence sampling. The result was subsequently improved in Ref. \cite{doi:10.1137/1.9781611975031.10} where a faster, exact algorithm was developed. The same authors further improved the runtime in Ref. \cite{clifford2020fasterclassicalbosonsampling}.
In general, the computational complexity for bosonic computations can be posed as an exponential scaling in terms of a measure of non-Gaussianity \cite{PhysRevResearch.3.033018}\cite{PhysRevLett.130.090602}. For GBS, the measurements constitute the non-Gaussian element. 
Phase-space based methods are useful for validating experiments \cite{Dellios2023}.

The state of the art in classical simulations of experimental GBS is based on tensor network methods and relies on the fact that the reduced density matrix of a Gaussian state can be diagonalized exactly \cite{Oh2024}. We explore a more straightforward approach, which can be further generalized to a larger set of multi-modal states, beyond Gaussian States, with application in e.g. non-linear boson sampling \cite{Spagnolo2023}.
Extending the input of a GBS protocol to non-Gaussian states does not increase the scaling of the complexity \cite{PhysRevA.109.052427}. It does, however, increase the resilience to noise in achieving a quantum advantage \cite{PhysRevA.106.042413}. Our method can be used as an approximate classical simulation of such proposals, with which to validate the quantum computation

In this work, we construct a Hamiltonian for which the resulting quantum state of the experiment is the unique ground state with energy zero. By writing the Hamiltonian in the Fock basis and solving the resulting variational problem with e.g. DMRG \cite{PhysRevLett.69.2863}, we obtain an approximate solution in polynomial time.
By construction, the variational energy is a direct measure of the simulation error, hence accuracy can be assessed on the fly, leading to a tradeoff between simulation time and accuracy as governed by the complexity of the variational ansatz.

One of the drawbacks of our approach, is that the discarding of high photon number Fock basis elements induces an error on the low photon number probabilities.
Therefore, we make use of Local Basis Optimization \cite{PhysRevLett.80.2661} (LBO), in order to achieve a higher cutoff at reduced computational complexity. For the case of GBS, we derive the exact optimal local basis, which additionally yields access to bounds on the truncation error.

For the general case it is not possible to derive the optimal local basis and we examine a complimentary approach to standard LBO techniques, that we denote parameterized LBO.
Simply stated, we allow our physical basis to have variational parameters which are optimized simultaneously with the variational state. Crucially, the local basis transformation is implemented analytically in the Hamiltonian, such that all computations are done in the optimized basis. Similar ideas can be found in ref. \cite{PhysRevLett.108.160401} \cite{PhysRevLett.102.150601}. We show that a better approximation of the ground state solution can be found efficiently using orders of magnitude fewer basis elements (i.e. a substantially smaller cutoff).

The article is structured as follows.
In section \ref{sec:theory} the main methodology of the paper is outlined along with an application to Gaussian Boson Sampling.
In section \ref{sec:methods}, local basis optimization is introduced and exactly solved in the case of GBS. Parameterized LBO is subsequently introduced and an optimization strategy in terms of tensor network methods is formulated. These optimization techniques are applied to a non-Gaussian sampling problem. Results are presented in sec. \ref{sec:results}.

\begin{figure*}
    \centering
    \includegraphics[width=1\textwidth,trim={0 0cm 0 0cm},clip]{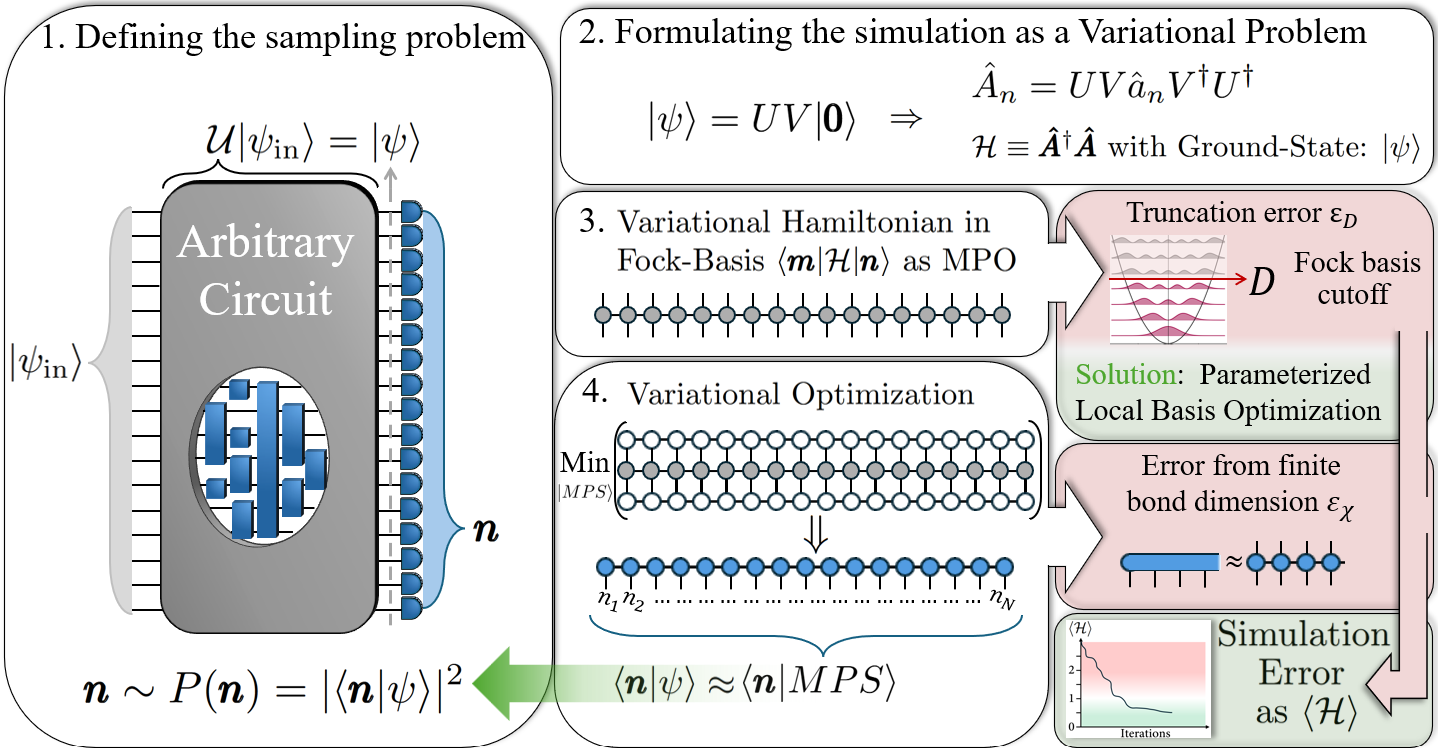}
    \caption{\label{fig:Method}
    Overview of our variational approach to solving Bosonic sampling problems.
    A Hamiltonian $\mathcal{H}$ can be constructed, for which the target state $|\psi\rangle$ is the unique ground state with energy zero. Once the Hamiltonian is written in the Fock Basis, standard variational methods can be used to obtain an approximation to the sampling probabilities $|\langle \pmb{n}|\psi\rangle|^2$. Furthermore, two approximation errors are identified, the magnitude of which can be assessed simply by observing how close the variational energy tend towards the true ground state energy; 0. In order to mitigate the error due to the Fock basis cutoff, we study the application of parameterized Local Basis Optimization techniques and non-linear optimization methods, the results of which are outlined in sec. \ref{sec:results}.
    }
\end{figure*}

\section{\label{sec:theory}Theory}
In this section, we present the main method of the paper and subsequently relate it to the relevant physical setting. For an illustrative overview, see figure \ref{fig:Method}. For further details on the Boson sampling problem see \cite{10.1117/1.AP.1.3.034001} \cite{PhysRevA.100.032326}. We set $\hbar=1$ throughout.

\subsection{Main Methodology}
Let us start by considering a noiseless quantum photonics experiment. A typical Boson sampling problem consists of evolving a known input state through a set of CV-gates such as beamsplitters and squeezing and measuring the photon number for each output mode. If we denote the pure state right before the PNR measurements by $|\psi\rangle$ then the probability of a specific outcome of the PNR measurements will follow the distribution:
\begin{equation}
    P(\pmb{n}) := P(n_1,n_2,\hdots,n_N) = \left|\left(\otimes_{k=1}^N \langle n_k| \right)|\psi\rangle\right|^2 =: |\langle \pmb{n}|\psi\rangle|^2
\end{equation}
where the vector $\pmb{n}$ refers to the number of photons in each mode and $|n_k\rangle$ refer to the $n$-th simple quantum harmonic oscillator state in the $k$-th mode.

When both the input state and the CV-gates are Gaussian, $|\psi\rangle$ will be a Gaussian state and the sampling problem is referred to as Gaussian Boson Sampling (sec. \ref{subsec:GBS}). The Gaussian state is completely characterized by a $2N\times 2N$ covariance matrix and a $2N$-dimensional displacement vector \cite{serafini2017quantum}. However, even then, calculating an overlap of the form $|\langle \pmb{n}|\psi\rangle|^2$, scales exponentially in $\sum_k n_k$ \cite{doi:10.1126/sciadv.abl9236}.

Instead, one can recover all probabilities at once by treating $\langle \pmb{n}|\psi\rangle$ as the solution to a variational problem for which one can employ tensor network methods, such as DMRG, to achieve an approximate solution in polynomial time. This formulation also allows us to treat non-Gaussian boson sampling problems.

\subsection{Variational Formulation}
We consider the noiseless case here. A discussion of the impact of noise is provided in section \ref{sec:SamplingWithNoise}.
Suppose we want to sample from a few body input state $|\psi_\text{in}\rangle$ that has been subject to the unitary gates of a circuit $\{U_i\}$.
\begin{equation}
    |\psi\rangle = \prod_{i\in \text{circuit}} U_i |\psi_\text{in}\rangle = U|\psi_\text{in}\rangle
\end{equation}
We will then make the crucial assumption that the initial state can be prepared from the vacuum, through a set of unitary gates $\prod_{i\in \text{state prep}}V_i = V$ and possibly PNR measurements of auxiliary modes.
This is in fact without loss of generality \cite{PhysRevResearch.3.043182}\cite{PhysRevLett.82.1784}. However, in general, an explicit construction of the circuit is not immediately available. A notable example that makes use of auxiliary modes, is when the input are Fock states. These can be heralded from two-mode squeezed vacuum states, e.g. as formulated in scattershot boson sampling \cite{PhysRevLett.113.100502}.

The method relies on the following simple observation:
\begin{equation}
\label{eq:CentralEquation}
    0 = \sum_n^N \hat{a}^\dagger_n\hat{a}_n|\pmb{0}\rangle = \sum_n^N\hat{A}_n^\dagger \hat{A}_n|\psi\rangle = \mathcal{H}|\psi\rangle
\end{equation}
where
\begin{equation}
    \hat{A}_n = U V \hat{a}_n V^\dagger U^\dagger
    \quad \quad
    |\psi\rangle = U V|\pmb{0}\rangle
\end{equation}
The state $|\psi\rangle$ will be the unique ground state of the positive semidefinite hamiltonian $\mathcal{H}$ with eigenvalue zero. The uniqueness comes from the preservation of the spectrum by the similarity transformations. Thus, $\mathcal{H}$ will also have a harmonic spectrum which is again evident from the conservation of the canonical commutation relations:
\begin{equation}
\label{eq:generalizedCCR}
    [\hat{A}_i,\hat{A}^\dagger_j] = \delta_{ij}
\end{equation}

Therefore, by obtaining the ground state of the matrix representation of the Hamiltonian in the Fock-basis $\langle \pmb{m}|\mathcal{H}|\pmb{n}\rangle$ one has acquired $\langle \pmb{n}|\psi\rangle$ for all PNR detection possibilities $\pmb{n}$ below some cutoff of the Fock-space and thus complete access to the sampled photon number distribution $|\langle\pmb{n}|\psi\rangle|^2$. Furthermore, the variational energy $\langle \mathcal{H}\rangle$ is directly related to the simulation error, as clarified in sec. \ref{sec:SimulationError}.

It remains to derive the Hamiltonian for experiments of interest and to answer the question of how to simulate real experiments in the presence of noise. This is done in the following sections.

\subsection{Deriving the Hamiltonian}
\subsubsection{\label{subsec:GBS}Gaussian Boson Sampling}
For Gaussian Boson sampling the vacuum is subject to single mode squeezing and a linear interferometer before being measured by photon number resolved detectors. 
For the following it will be natural to first rotate the $\hat{a}$ and $\hat{a}^\dagger$ into the $\hat{X}$ and $\hat{P}$ quadrature operators:
\begin{align}
\begin{pmatrix}
    \hat{a}_k \\
    \hat{a}_k^\dagger
\end{pmatrix}
&=\frac{1}{\sqrt{2}}
\begin{pmatrix}
    1 & i \\
    1 & -i
\end{pmatrix}
\begin{pmatrix}
    \hat{X}_k \\
    \hat{P}_k
\end{pmatrix}
\\
    \mathcal{H}_{|0\rangle} &= \sum_k\hat{a}_k^\dagger\hat{a}_k =
    \frac12
    \begin{pmatrix}
        \hat{\pmb{X}} \\
        \hat{\pmb{P}}
    \end{pmatrix}^T
    \begin{pmatrix}
        \hat{\pmb{X}} \\
        \hat{\pmb{P}}
    \end{pmatrix}
    - \frac{N}{2}
\end{align}
The squeezing operation simply rescales the quadrature operators in an inverse fashion to what is performed on the state:
\begin{equation}
    \mathcal{H}_{V^\dagger|0\rangle} = \sum_k V^\dagger \hat{a}_k^\dagger \hat{a}_kV = 
    \frac12
    \begin{pmatrix}
        \hat{\pmb{X}} \\
        \hat{\pmb{P}}
    \end{pmatrix}^T
    D^{-2}
    \begin{pmatrix}
        \hat{\pmb{X}} \\
        \hat{\pmb{P}}
    \end{pmatrix}
    - \frac{N}{2}
\end{equation}
Where $D$ is a diagonal matrix: $D =\text{diag} 
\begin{pmatrix}
    e^{-r_1} &\hdots & e^{-r_N} & e^{r_1} & \hdots & e^{r_N}
\end{pmatrix}$
with $\{r_i\}\in\mathbb{R}$ specifying the squeezing amount of each mode in the Gaussian state. The transformation under the passive Gaussian transformation rotates the quadrature operators by the orthogonal matrix $O$:
\begin{equation}
    \mathcal{H} = \sum_k U^\dagger V^\dagger \hat{a}_k^\dagger \hat{a}_kVU
    =\frac12
    \begin{pmatrix}
        \hat{\pmb{X}} \\
        \hat{\pmb{P}}
    \end{pmatrix}^T
    O^TD^{-2}O
    \begin{pmatrix}
        \hat{\pmb{X}} \\
        \hat{\pmb{P}}
    \end{pmatrix}
     - \frac{N}{2}
\end{equation}
We can express the Hamiltonian in terms of the covariance matrix of the Gaussian state $V=\frac12 OD^2O^T$ \cite{serafini2017quantumCov} namely:
\begin{equation}
\label{eq:GBSHamiltonian}
    \mathcal{H} = 
    \begin{pmatrix}
        \hat{\pmb{X}}\\
        \hat{\pmb{P}}
    \end{pmatrix}^T
    \frac{V^{-1}}{4}
    \begin{pmatrix}
        \hat{\pmb{X}}\\
        \hat{\pmb{P}}
    \end{pmatrix} - \frac{N}{2}
\end{equation}
The ground state of such an n-dimensional anisotropic oscillator in a uniform magnetic field \cite{osti_4699813} is exactly the Gaussian state in question. Displacements are easily accounted for in the simulation (they constitute single mode gates) and are therefore neglected here.

\subsubsection{\label{subsec:VarHamilBeyondGBS}Beyond GBS: fully connected $CZ$-gate.}
GBS poses a benchmark problem; a setting for which we can compare our method to other exact methods. In order to highlight the flexibility of our approach and to showcase our novel methods for solving the variational problem, we also consider a (artificial) highly non-Gaussian circuit that would be not so straight forward by other means. That is, a Gaussian circuit on which a completely connected phase gate is applied in between the Gaussian unitary and the PNR measurements. The reason for the full connectivity is to ensure that we can't just solve for the Gaussian state and apply unitary gates to our representation of the state (this would for example be a viable approach if we only applied, the otherwise non-Gaussian, cubic phase gate to each output mode).
Specifically we consider:
\begin{equation}
\label{eq:GlobalCZgate}
    |\psi\rangle = \exp{\left(-i\kappa\prod_k^N\hat{X}_k\right)}|\psi_\text{Gaussian}\rangle
\end{equation}
which leaves $\hat{X}$ invariant but transforms $\hat{P}$ as:
\begin{equation}
    \hat{P}_i \rightarrow \hat{P}_i + \kappa \prod_{n\neq i} \hat{X}_n
\end{equation}
such that if we define a "non-local" vector of operators: 
\begin{equation}
    (\pmb{\hat{\xi}})_i=\kappa \prod_{n\neq i} \hat{X}_n
\end{equation}
The variational Hamiltonian for this example, is simply:
\begin{equation}
\label{eq:nonGBSHamiltonian}
    \mathcal{H} = 
    \begin{pmatrix}
        \hat{\pmb{X}}-\pmb{\mu}\\
        \hat{\pmb{P}} + \hat{\pmb{\xi}} - \pmb{\nu}
    \end{pmatrix}^T
    \frac{V^{-1}}{4}
    \begin{pmatrix}
        \hat{\pmb{X}}-\pmb{\mu}\\
        \hat{\pmb{P}} + \hat{\pmb{\xi}} - \pmb{\nu}
    \end{pmatrix} - \frac{N}{2}
\end{equation}
The presence of $\hat{\pmb{\xi}}$ makes the Hamiltonian non-Gaussian when we consider a circuit with $N>2$.

\subsection{Matrix Product States}
In order to numerically treat many body quantum systems, it is necessary to restrict the state representation to only a part of the Hilbert space. Matrix Product States (MPSs) provide a succinct description of the physically relevant states, with direct access to truncation of the entanglement spectrum between the constituent Hilbert spaces of the N-body Hilbert space \cite{biamonte2020lecturesquantumtensornetworks}\cite{Bridgeman_2017}\cite{annurev:/content/journals/10.1146/annurev-conmatphys-040721-022705}.

The Matrix Product State ansatz is a statement about the tensor $c_{n_1,n_2,\hdots}$ containing the coefficients of the pure state in a given basis:
\begin{equation}
\label{eq:GenericState}
    |\psi\rangle = \sum_{\pmb{n}} c_{\pmb{n}}|n_1\rangle \otimes |n_2\rangle \otimes \hdots \otimes |n_N\rangle
\end{equation}
Namely, that it can be written as a product of matrices:
\begin{equation}
\label{eq:MPSAnsatz}
    c_{\pmb{n}} = (\pmb{M}^{(1)}_{n_1})^T M^{(2)}_{n_2}\hdots M^{(N-1)}_{n_{N-1}}\pmb{M}^{(N)}_{n_N}
\end{equation}
Here $\pmb{M}^{(1)}$ and $\pmb{M}^{(N)}$ are vectors because we choose open boundary conditions for algorithmic reasons.
Diagramatically, eq. \ref{eq:MPSAnsatz} looks like (for $N=5$):
\begin{equation}
    \adjincludegraphics[valign=c,width=0.4\linewidth]{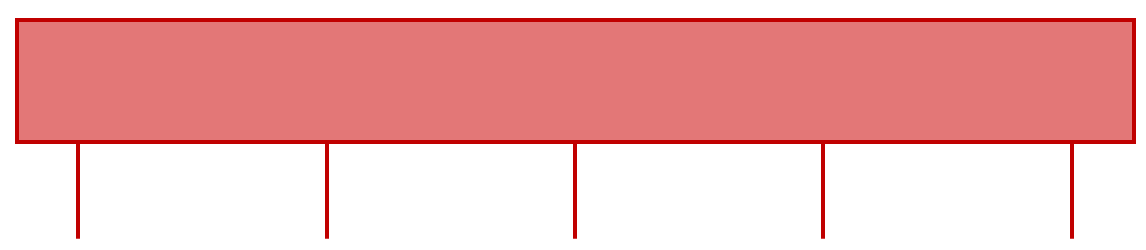} = \adjincludegraphics[valign=c,width=0.4\linewidth]{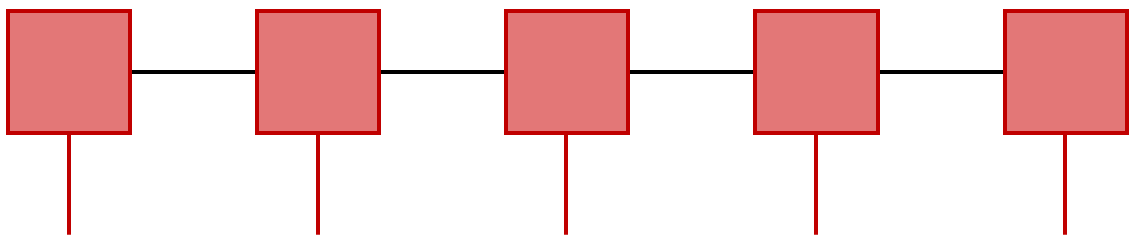}
\end{equation}
where the legs denote the indices of the corresponding tensor and connected legs denote summation over that index. The matrices $\{M^{(i)}_{n_i}\}$ are of dimension $\chi\times\chi$ where the \textit{bond dimension} $\chi$ reflects the entanglement across the specific cut of the chain. Systems with extensive entanglement require a large bond dimension to be represented faithfully as an MPS, while low entanglement states can be efficiently described with small bond dimension. Specifically, whether a state permits an efficient MPS representation can be diagnosed from the scaling of its Rényi entropy \cite{PhysRevLett.100.030504}\cite{PhysRevB.73.094423}.

Representation of the state as an MPS, allows for powerful variational methods, such as the DMRG-algorithm \cite{PhysRevLett.69.2863}, which is able to optimize the full state by optimizing the single-site matrices one by one in a sequential manner \cite{SCHOLLWOCK201196}.
Our parameterized local basis optimization approach (sec. \ref{subsec:pLBO}) builds upon the MPS structure.

\subsection{\label{sec:SimulationError}Simulation Errors}
In order to represent the bosonic operators numerically, we have to introduce a cutoff of the infinite dimensional space. This cutoff together with the truncation of the ansatz space to MPS with finite bond-dimension: 
\begin{equation*}
    \text{Fock basis cutoff: } D, \quad\text{MPS bond dimension: } \chi
\end{equation*}
constitute the two hyper-parameters of the variational problem, and together fully control the simulation error.\\

One of the key properties of our variational method is that the ground state energy is zero in the continuum. Therefore, any truncation or approximation error will increase the variational energy $\langle \mathcal{H}\rangle$ above zero.
In other words:
\begin{equation}
\label{eq:MagOfSimError}
    \text{"Magnitude of Simulation Error"} \equiv \langle \mathcal{H} \rangle \geq 0
\end{equation}
A concrete manifestation of this relation can be put forward by exploiting the unit-sized harmonic spectrum of the hamiltonian:
\begin{equation}
    \mathcal{H} = 0\cdot|\psi\rangle\langle\psi| + \sum_{n=1}nP_n
\end{equation}
where $P_n = \sum_i |n_i\rangle\langle n_i|$ is a projection onto the subspace spanned by the degenerate eigen-states with energy $n$. Consider then the expectation value of $\mathcal{H}$ wrt. the variational state:
\begin{equation}
    \langle\mathcal{H}\rangle = \sum_{n=1} n\langle P_n\rangle \geq\sum_{n=1}\langle P_n\rangle = 1 - F
\end{equation}
where $F$ is the absolute square of the overlap between the variational state and $|\psi\rangle$ and constitutes the fidelity of the simulation. Thus $F\geq 1-\langle\mathcal{H}\rangle$ which is confirmed in sec. \ref{sec:results} (fig. \ref{fig:GBSErrors}c). If $\langle\mathcal{H}\rangle \geq 1$ the bound is rendered uninformative. However, in principle, the probability distribution $\langle P_n\rangle$ and thereby $F$ can be deduced by considering higher order moments of the variational energy $\langle \mathcal{H}^k\rangle$. This is addressed in the SM sec. 1 together with any concerns towards complications from truncation effects, which will now be outlined.

\subsubsection{Errors from the cutoff $D$}
Truncating the infinite-dimensional operators introduces an error in the ground state eigenvalue problem.
Let $P$ be the projection
\begin{equation}
\label{eq:ProjectionOperator}
P=\bigotimes_i^N\left(\sum_{n_i=0}^{D-1}|n_i\rangle\langle n_i|\right)
\end{equation}
and $Q = \mathbbm{1} - P$ the complement. Then it becomes apparent that the eigendecomposition of the naively truncated problem does not in general completely align with the eigendecomposition of the original problem. Namely, we are interested in $P|\psi\rangle$ where
\begin{equation}
    \mathcal{H}|\psi\rangle = 0
\end{equation}
However, from insertion of the identity we see that:
\begin{equation}
\label{eq:ProjectedEigenProblem}
    P\mathcal{H}P|\psi\rangle = - P\mathcal{H}Q|\psi\rangle
\end{equation}
So unless $|\psi\rangle$ is completely contained in the projected space (i.e. $Q|\psi\rangle=0$) the ground state of $P\mathcal{H}P$ will be misaligned with $P|\psi\rangle$.

The exact effective Hamiltonian, for which the ground state will align with $P|\psi\rangle$, is found by also applying $Q$ from the left on eq. \ref{eq:ProjectedEigenProblem} and solving for $P|\psi\rangle$ \cite{PhysRevD.91.085011}:
\begin{equation}
\label{eq:EffectiveHamiltonian}
    (P\mathcal{H}P - P\mathcal{H}Q(Q\mathcal{H}Q)^{-1}Q\mathcal{H}P)P|\psi\rangle=0
\end{equation}
In general, the exact effective Hamiltonian is not available, and one would resort to perturbation theory or targeted unitary transformations to achieve an approximate effective description \cite{PhysRevB.101.014302}. In this work we simply consider $P\mathcal{H}P$, incurring the error:
\begin{equation}
\label{eq:epsilonD}
     \varepsilon_D := \big|\big| |\psi\rangle - P|\psi\rangle \big|\big|^2 = 1 - \langle \psi |P|\psi\rangle
\end{equation}
$\varepsilon_D$ captures the rotation induced on the ground state when going from the full description to the naively truncated description. For GBS $\varepsilon_D$ can be directly assessed as shown in sec. \ref{subsec:LBO}.

\subsubsection{Assessing the error from finite bond dimension}
Any quantum state can be cast as an MPS. However, in general, the bond dimension will have to grow exponentially with system size. If instead the bond dimension is capped at a constant value, an approximation error is introduced.

In the case of GBS the compression of the Gaussian state to its MPS form is based on sequential singular value decompositions \cite{PhysRevLett.91.147902}\cite{Oh2024}. Therefore, we denote the compressed state $|SVD\rangle$. The singular values themselves inform about the error introduced by a finite bond dimension $\chi$, which we denote $\varepsilon_\chi$ (see SM sec. 3 for a formal description \cite{supp}). Specifically, they provide an upper bound on $\varepsilon_\chi$, while mitigating the complications of the infinite dimensional Hilbert space, and therefore we use these as a surrogate \cite{OSELEDETS201070}\cite{doi:10.1137/090752286}\cite{doi:10.1137/S0895479896305696}:
\begin{equation}
\label{eq:SurrogateEpsChi}
    \varepsilon_\chi \sim \sum_{n=1}^{N-1}\left(1 - \sum_{k\leq\chi_n}\left(\sigma^{(n)}_k\right)^2\right)
\end{equation}
$\sigma^{(n)}$ are the singular values and are equal to the Schmidt coefficients when bipartitioning the state into $\{1,\hdots,n\}$ and $\{n+1,\hdots,N\}$.

Outside GBS, it is impractical to obtain the compressed state through sequential SVDs of the full state. Instead, the MPS approximation $|MPS\rangle$ to the ground state $|GS\rangle$ of the Hamiltonian $P\mathcal{H}P$ is obtained through variational methods. The error induced by imposing an MPS ansatz on the variational state $|MPS\rangle$, can generally be estimated by its variance wrt. the Hamiltonian \cite{McCulloch_2007}:
\begin{equation}
    \sigma_{\mathcal{H}} = \sqrt{\langle (P\mathcal{H}P)^2\rangle - \langle P\mathcal{H}P\rangle^2}
\end{equation}
The compression error $\varepsilon_\chi'\sim\sigma_\mathcal{H}$ is the degree to which $|MPS\rangle$ remains an eigenstate of the Hamiltonian.
Schematically, the relation between the relevant states in terms of the errors introduced is:
\begin{align}
    &\quad \: \: \: |GS\rangle \xrightarrow[]{\varepsilon_\chi'} |MPS\rangle \nonumber\\
    &\nearrow_{\varepsilon_D} \nonumber\\
    |\psi\rangle& \label{SchematicStates}\\
    &\searrow^{\varepsilon_\chi} \nonumber\\
    &\quad \: \: \: |SVD\rangle \nonumber
\end{align}
with $|\psi\rangle$, $|SVD\rangle$, $|GS\rangle$ and $|MPS\rangle$ being the target-, compressed-, ground- and variational state, respectively.

In conclusion, by introducing finite-dimensional matrix representations of the bosonic operators and using Tensor Networks for the variational ansätze, the simulation becomes approximate while being controlled by the parameters $D$ and $\chi$. In order to work towards no error ($\langle\mathcal{H}\rangle = 0$, $D$ and $\chi$ should be increased and the value of $\sigma_\mathcal{H}$ informs which to increase, since $\sigma_\mathcal{H}\approx 0$ means that the current $\chi$ is adequate.

\subsection{\label{sec:SamplingWithNoise}Sampling in Presence of Noise}
For experimental GBS, the proper description will be in terms of a mixed state and it is not immediately clear how to apply our method which derived the variational Hamiltonian from a pure state. However, for mixed state GBS, it is possible to divide the sampling problem into two parts \cite{PRXQuantum.3.010306}\cite{Oh2024}. 
Namely, a pure state GBS problem followed by a classical sampling problem, which can be done efficiently.
Specifically, one decomposes the mixed state covariance matrix into two using semi-definite programming:
\begin{equation}
\label{eq:CovarianceDecomposition}
    V = Q + C
\end{equation}
$Q$ constitutes a pure state covariance matrix and contains all the entanglement information and the exponentially hard computational complexity. $C$ is the covariance matrix of a classical zero-mean Gaussian distribution of random displacements, which encapsulates all the noise of the experiment. After obtaining the pure state defined by $Q$, random displacement operations are applied onto this state and a PNR measurement is sampled from the resulting state.
These samples will follow the distribution defined by the noisy GBS problem.
The computationally hard part of this problem is obtaining the pure state amplitudes $\langle \pmb{n}|\psi\rangle$, which can be done with our method.

In general, noise in the form of photon loss, present throughout the circuit, can be accounted for through the use of auxiliary degress of freedom. For each lossy mode, entanglement is transferred through a beamsplitter to an auxiliary mode that is subsequently traced out.
Thus, one can obtain the pure state $\langle \pmb{n}|\psi\rangle$ of the extended circuit and after tracing out the auxiliary degrees of freedom, obtain the density matrix of the noisy sampling problem.

\section{\label{sec:methods}Methods}
The variational formulation, as described in sec. \ref{sec:theory}, is agnostic to the choice of optimization procedure for recovering the ground state. In this paper, we employ two different optimization strategies, that are both based on tensor networks, but where one is linear and the other non-linear.

For the Gaussian case, we will introduce Local Basis Optimization and solve it exactly, such that GBS can be simulated in the optimal local basis by DMRG. For non-Gaussian sampling problems we introduce the parameterized LBO method and an accompanying non-linear optimization strategy.

\subsection{\label{subsec:LBO}Optimal Local Basis and Cutoff Error Bounds}
It is of central importance to reduce the error associated to the truncation of the bosonic operators. To this end Local Basis Optimization seeks to reduce the physical dimension $D$ of the local Hilbert space by introducing a new local basis of smaller dimension $d\ll D$ \cite{PhysRevB.62.R747}. The optimal local basis depends on the state that is being considered and is given by the eigenvectors of the reduced density matrix:
\begin{equation}
\label{eq:ReducedDensityMatrix}
    \rho_i = \text{Tr}_{k \neq i}(|\psi\rangle\langle\psi|)
\end{equation}
The error induced by decreasing the physical dimension is related to the eigenvalues of the reduced density matrix and is minimized by keeping the eigenvectors corresponding to the largest eigenvalues.

LBO can also be viewed as a compression of the physical index by SVDs:
\begin{equation}
\label{eq:SVDPhysicalDimension}
    \adjincludegraphics[valign=c,width=0.3\linewidth]{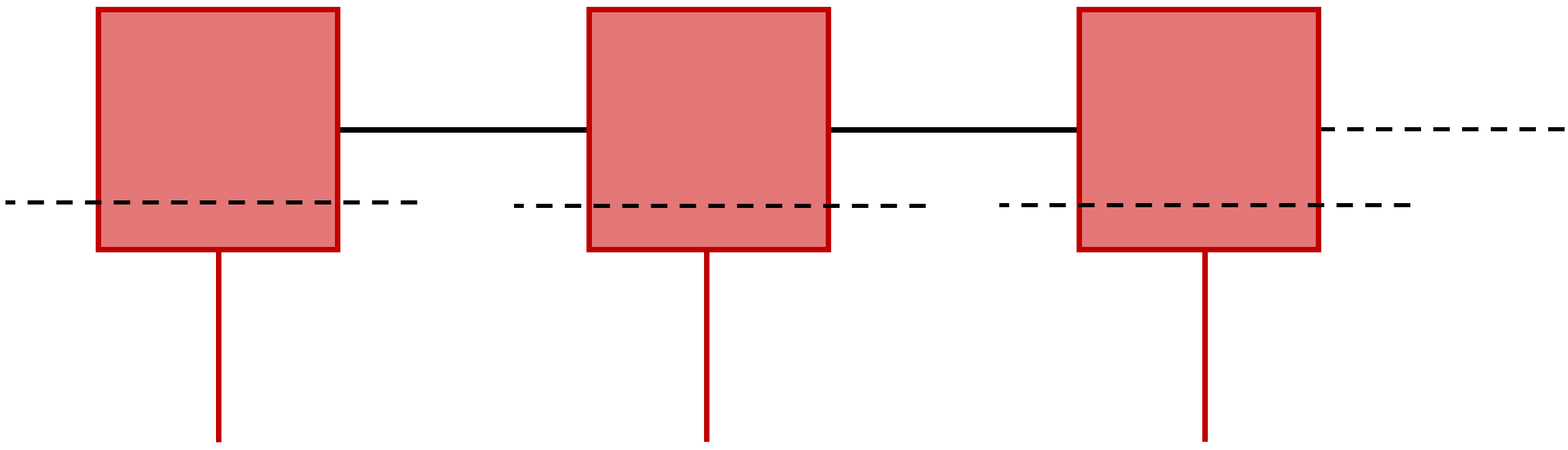} = \adjincludegraphics[valign=c,width=0.3\linewidth]{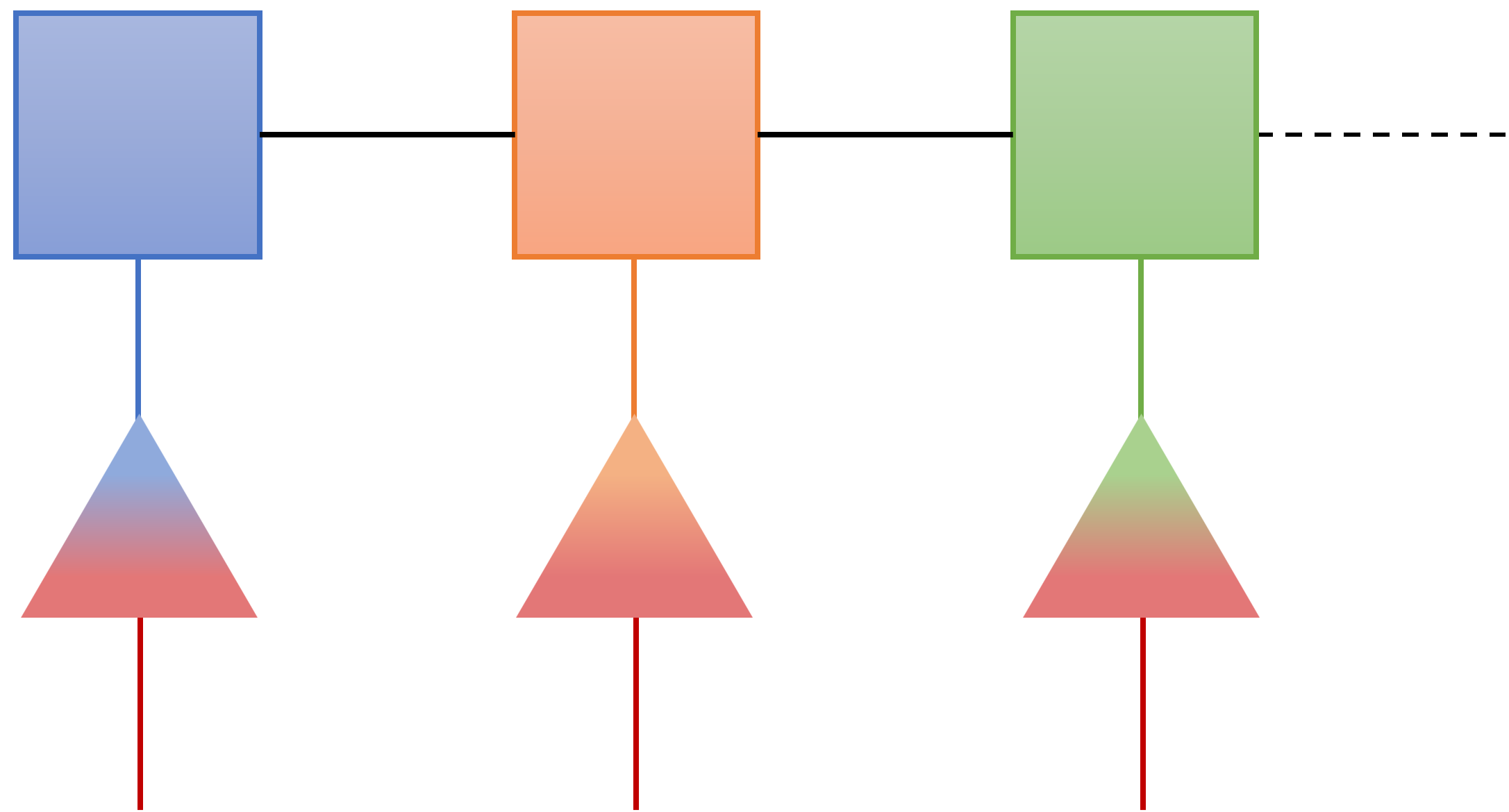}
\end{equation}
where each color denote their own basis and the red color denotes the basis comprised of photon number eigenstates. The Hamiltonian can be cast in the optimal local basis:
\begin{equation}
    \adjincludegraphics[valign=c,width=0.3\linewidth]{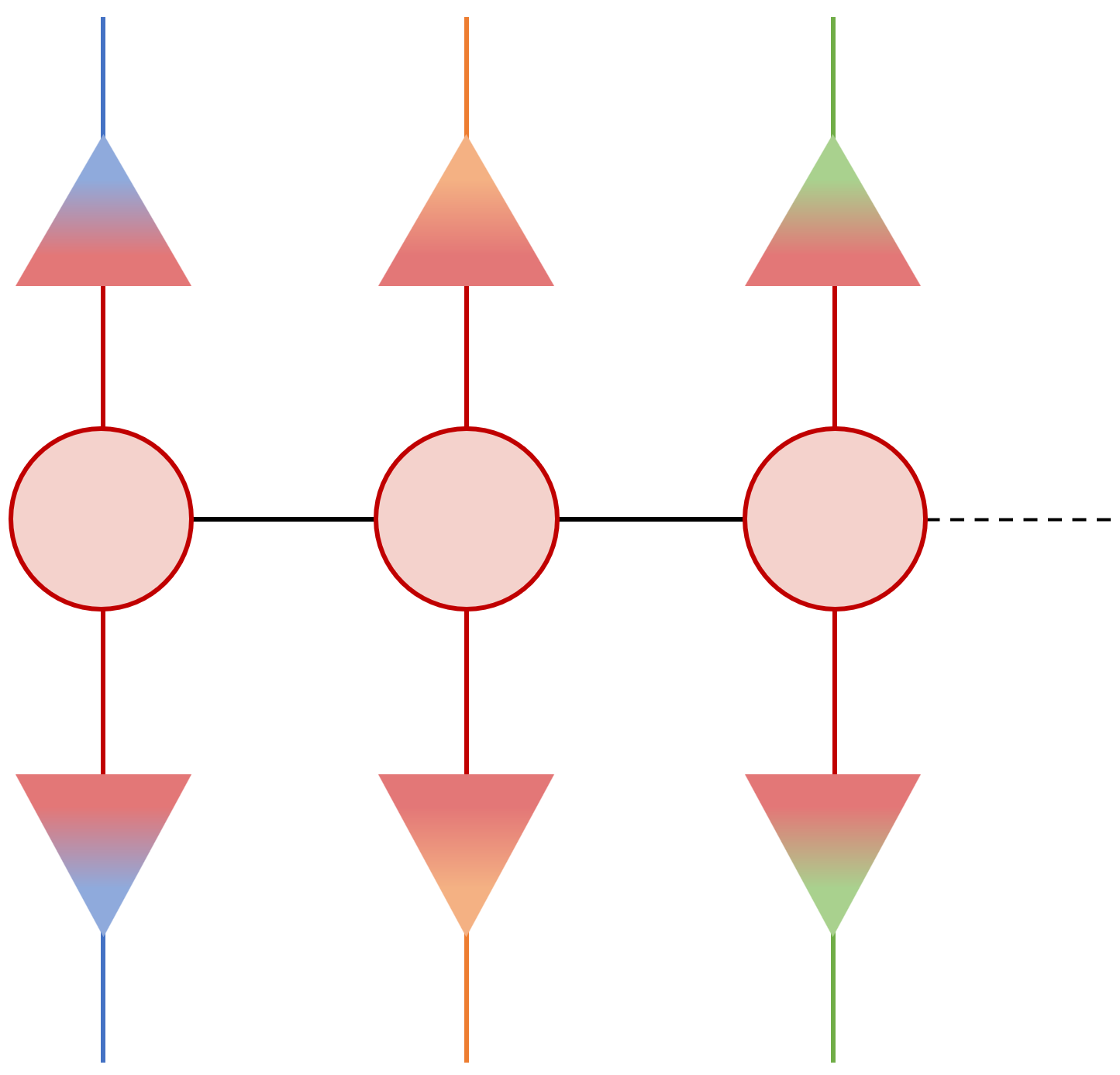} = \adjincludegraphics[valign=c,width=0.3\linewidth]{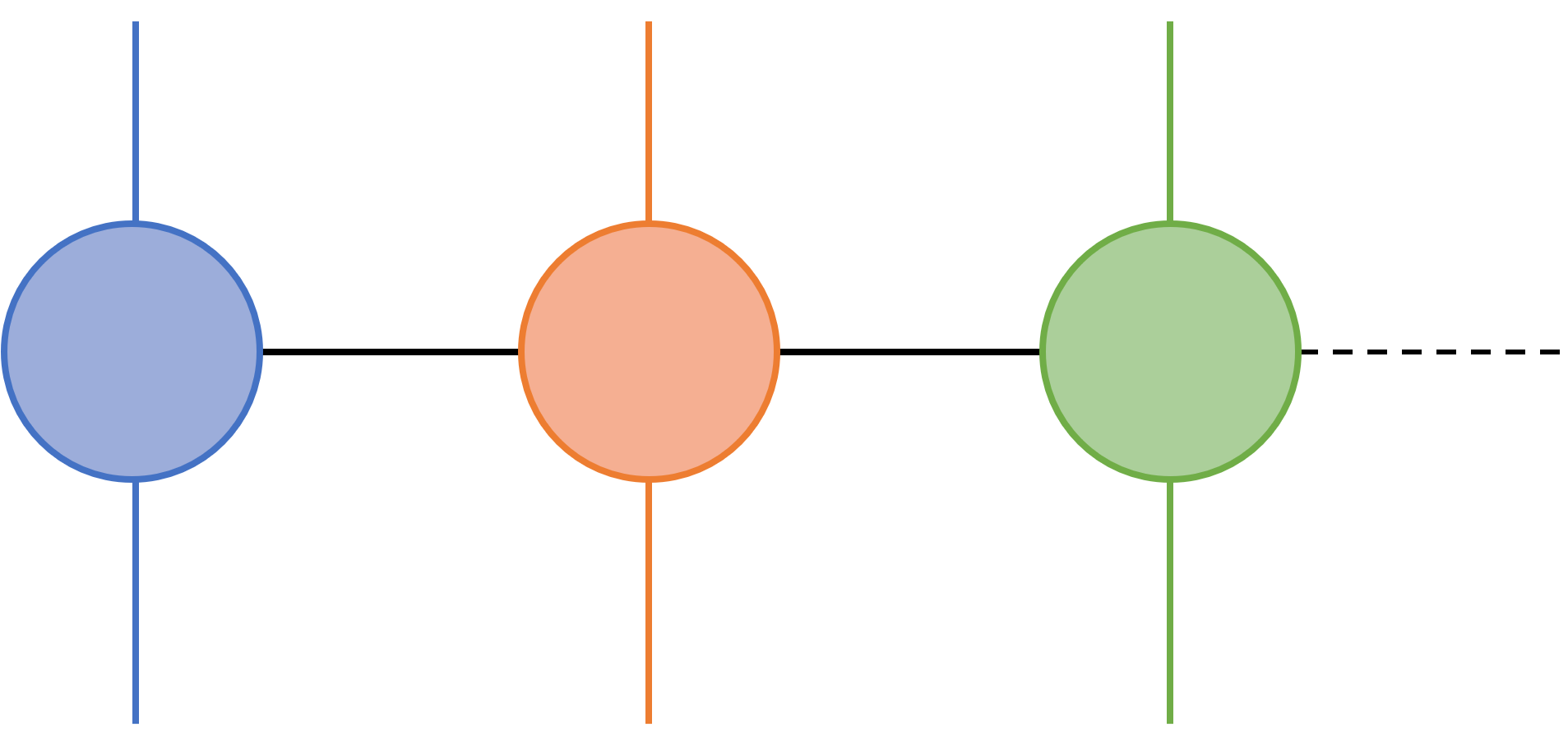}
\end{equation}
thus reducing the numerical effort for subsequent calculations.

Usually the target state is not known beforehand and LBO has to be carried out simultaneously with the variational optimization in a self-consistent manner \cite{PhysRevLett.80.2661}\cite{PhysRevB.92.241106}. However, in our setting, the target state is known and specifically for Gaussian states, it is possible to derive the optimal local basis analytically.

\subsubsection{Optimal Local Basis for Gaussian States}
The reduced density matrix of a Gaussian state is found by integrating out all the modes $k$ different from $i$ and is itself a Gaussian state.
The Gaussian density matrix is diagonalized as \cite{serafini2017quantum}:
\begin{equation}
\label{eq:diagReducedGaussianDM}
    \rho_i = \sum_{m_i=0}\frac{\bar{n}_i^{m_i}}{(\bar{n}_i+1)^{m_i+1}}\hat{S}_i^\dagger|m_i\rangle\langle m_i|\hat{S}_i
\end{equation}
with $\bar{n}_i=\langle\hat{n}_i\rangle_G$ being the mean number of photons and $|m_i\rangle$ the photon number eigenstates. 
The Hamiltonian is rotated to the basis spanned by $\hat{S}^\dagger_i|m_i\rangle$ by transforming under $\hat{S}_i$:
\begin{equation}
\label{eq:OLB}
    \mathcal{H} \rightarrow \hat{S}_i \mathcal{H} \hat{S}_i^\dagger
\end{equation}
The action of the quadratic unitary $\hat{S}_i$ on the quadrature operators is:
\begin{equation}
    \hat{S}_i
    \begin{pmatrix}
        \hat{X}_i \\
        \hat{P}_i
    \end{pmatrix}
    \hat{S}_i^\dagger = S_i
    \begin{pmatrix}
        \hat{X}_i \\
        \hat{P}_i
    \end{pmatrix}
\end{equation}
Where the matrix representation $S_i$ is found by Williamson decomposition \cite{a6cda7d0-29df-361a-a64d-8dd6c762bce6} of the covariance matrix of the reduced Gaussian density matrix $V_i$ \cite{houde2024matrixdecompositionsquantumoptics}:
\begin{equation}
\label{eq:WilliamsonDecomposition}
    V_i = S_i
    \begin{pmatrix}
       \hbar \left(\bar{n}_i + \frac12\right) & 0 \\
       0 & \hbar \left(\bar{n}_i + \frac12\right)
    \end{pmatrix}
    S_i^T
\end{equation}
Notice from eq. \ref{eq:GBSHamiltonian} and eq. \ref{eq:WilliamsonDecomposition} that eq. \ref{eq:OLB} has the effect of diagonalizing the submatrix for mode $i$. This procedure is carried out for each mode.

Solving the variational problem in the optimal local basis $\prod_i \hat{S}_i^\dagger|\pmb{m}\rangle$ does not grant immediate access to the wanted Fock state overlaps $\langle \pmb{n}|\psi\rangle$. However, the fact that the basis transformation was local, means that it is a simple matter to apply the matrix representation of the inverse transformation:
\begin{equation}
\label{eq:InverseBasisTransformation}
    \langle n_i |\psi\rangle = \sum_{m_i}^{d_i} \langle n_i |\hat{S}_i^\dagger|m_i\rangle\langle m_i|\psi_\text{LBO}\rangle
\end{equation}
where $\langle m_i|\psi_\text{LBO}\rangle$ is to be understood as the result obtained when solving the variational problem defined by the local basis optimized Hamiltonian (eq. \ref{eq:OLB}). Note, that while the representation has a cutoff $d_i$ we can calculate the overlap for any $\langle n_i|$.  This results in the  notion of an effective cutoff for which the matrix $V \in \mathbb{C}^{D\times d}$ with $V_{nm} = \langle n_i|\hat{S}^\dagger_i|m_i\rangle$ is an isometry (up to machine precision):
\begin{equation}
\label{eq:Effectivecutoff}
    \text{"Effective cutoff"} = \text{minimal } D \text{ s.t. } V^\dagger V = \mathbbm{1}_{d\times d}
\end{equation}

\subsubsection{Truncation Error from Singular Values}
\begin{figure}[h]
    \centering
    \includegraphics[width=1\linewidth]{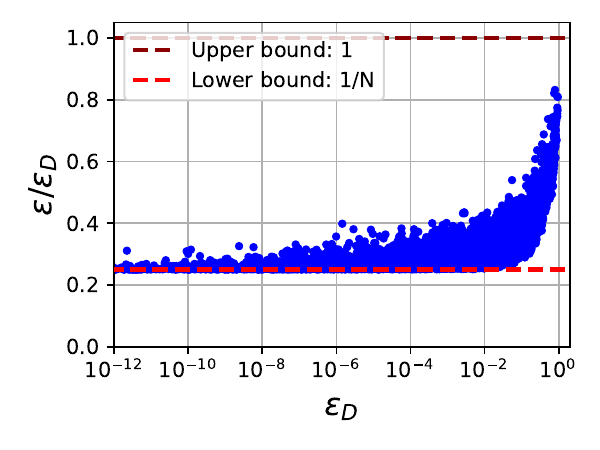}
    \caption{\label{fig:TruncationErrorBounds}
    This shows the bound as: $\frac{1}{N} \leq \frac{\varepsilon}{1-\langle \psi | P | \psi \rangle} \leq 1$ with $N=4$. The data suggests that for low enough truncation errors, the spectral weight can be assessed by: $\varepsilon_D \approx N\varepsilon$
}
\end{figure}

Besides obtaining the optimal local basis from the unitaries in eq. \ref{eq:diagReducedGaussianDM}, having access to all the singular values 
\begin{equation}
    \left(\sigma ^{(i)}_k\right)^2 \equiv \frac{\bar{n}_i^{k}}{(\bar{n}_i+1)^{k+1}}
\end{equation}
is also beneficial.
Similar to eq. \ref{eq:SurrogateEpsChi}, there is an inequality associated to the SVD compression of eq. \ref{eq:SVDPhysicalDimension} and it is a simple matter to prove (see SM sec. 4 \cite{supp}) that the spectral weight left after projection is bounded by the sum of the remaining singular values:
\begin{equation}
\label{eq:SpecWfromSDsums}
    N\varepsilon \geq \underbrace{1 - \langle \psi | P | \psi \rangle}_{\varepsilon_D} \geq \varepsilon
\end{equation}
where $0\leq \varepsilon \leq 1$ is defined as:
\begin{equation}
    \varepsilon := 1- \frac{1}{N}\sum_{i=1}^N\sum_{k\leq d_i} \left(\sigma_k^{(i)}\right)^2
\end{equation}
Empirically we find that the upper limit saturates as $\varepsilon$ approaches zero (see figure \ref{fig:TruncationErrorBounds}).

\subsection{\label{subsec:pLBO}Parameterized Local Basis Optimization}
Since we are unable to derive the equivalent of eq. \ref{eq:diagReducedGaussianDM} for a generic non-Gaussian state, we have to resort to approximate methods for the local basis optimization.
Standard techniques \cite{PhysRevLett.108.160401} consists of variationally optimizing $d$ orthonormal linear combinations of the $D\gg d$ lowest lying photon number states, which then serves as the basis in which the MPS is optimized. The local basis can be improved by simultaneously shifting the photon number basis by the expectation value of $\hat{X}_i$ during the optimization. The displacement is most aptly implemented directly in the matrix representation of the operators. These three steps are iterated upon until convergence and the next site is optimized in a similar fashion.

Incorporating more general local transformations other than displacement will enhance the representational power of the basis. By utilizing non-linear optimization methods, the associated basis parameters can be learned together with the basis, circumventing the need for sequential optimizations. We study the effects of this approach without the $(D\times d)$ dimensional linear basis transformation, however, the two methods can be combined, as we show in the supplemental material sec. 6 \cite{supp}. 

In choosing a parameterization for the local basis transformation, it is important to consider how the inverse transformation is obtained (eq. \ref{eq:InverseBasisTransformation}). To this end, we consider the operations included in the python module StrawberryFields \cite{Killoran2019strawberryfields}\cite{Bromley_2020}. Through the use of this library, we can obtain the matrix representation of the transformation from the local basis to the original Fock basis.

\begin{figure}
    \centering
    \includegraphics[width=1\linewidth]{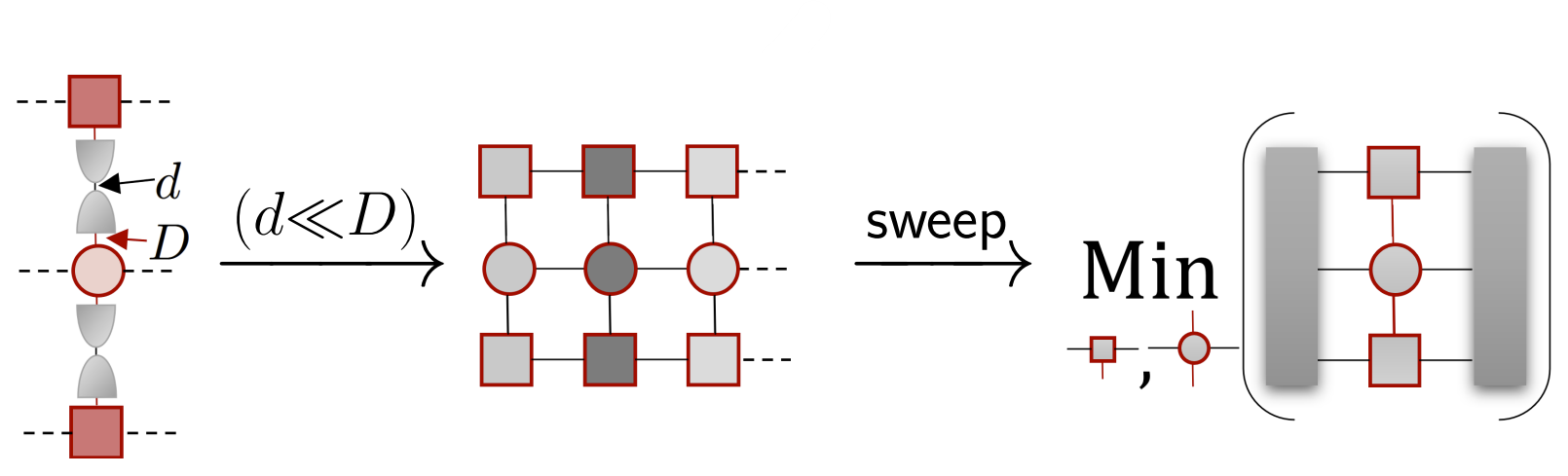}
    \caption{\label{fig:schematic_algo_plbo}Schematic of our variational algorithm that utilizes parameterized local basis optimization. It works much like DMRG, in that it defines an effective Hamiltonian for each individual site and performs an optimization for that site. The difference is that the MPO also contains variational parameters which are optimized using Gradient Descent algorithms due to the non-linearity introduced by eq. \ref{eq:SymplecticBasisTransformation} }
\end{figure}

There are multiple ways of composing a local basis unitary from the single site gates. The choice will affect the implementability and the size of the space the variational ansatz exists in. We choose a simple composition which make use of each gate once to become the following parameterized unitary:
\begin{equation}
\label{eq:pLBOUnitary}
    \mathcal{U}^\dagger = K^\dagger(\kappa) P_3^\dagger(\gamma) P_2^\dagger(s) R^\dagger(\theta) S^\dagger(re^{i\phi}) D^\dagger(\underbrace{\alpha_x + i\alpha_p}_{\alpha})
\end{equation}
The gates are from left to right the Kerr, Cubic phase, quadratic phase, rotation, squeezing and displacement gate.
The additional variational parameters are $\{\alpha_x,\alpha_p,r,\phi,\theta,s,\gamma,\kappa\}$, all of which are $\in\mathbb{R}$. Each mode has its own set of tuneable basis parameters. 
The action of $\mathcal{U}$ on the constituents of the Hamiltonian is:
\begin{equation}
\label{eq:SymplecticBasisTransformation}
\begin{aligned}
    \mathcal{U}^\dagger \hat{a} \mathcal{U} &= \hat{a}e^{i\kappa(2\hat{n}-1)}\left(\left(1 + i\frac{s}{2}\right)e^{i\theta}\cosh r + i \frac{s}{2}e^{i(\phi-\theta)}\sinh r\right)\\
    &- e^{-i\kappa(2\hat{n}-1)}\hat{a}^\dagger \left(\left(1 - i \frac{s}{2}\right)e^{i(\phi-\theta)}\sinh r - i \frac{s}{2}e^{i\theta}\cosh r\right) \\ 
    &+ i\frac{\gamma}{2}(e^{i\theta}\cosh r + e^{i(\phi-\theta)}\sinh r) \\
    &\quad\quad\times(2\hat{n}+1+\hat{a}\hat{a}e^{i4\kappa(\hat{n}-1)}+e^{-i4\kappa(\hat{n}-1)}\hat{a}^\dagger \hat{a}^\dagger) + \alpha
\end{aligned}
\end{equation}
The matrix representation of the transformed operator is easily obtained, as outlined in the SM sec. 5 \cite{supp}. For reasons discussed in sec. \ref{sec:discussion} we adopt an optimization strategy inspired by single site DMRG. We only optimize the variational parameters related to a single site before moving on to the next, thereby optimizing the full state and the local basis in a sweeping manner, as shown in fig. \ref{fig:schematic_algo_plbo}.

\section{\label{sec:results}Results}
\subsection{Gaussian Boson Sampling}

\begin{figure*}
\centering
\subcaptionbox{}{\raisebox{+.09\height}{\includegraphics[width=0.31\textwidth,trim={0.26cm 0.28cm 0.3cm 0.25cm},clip]{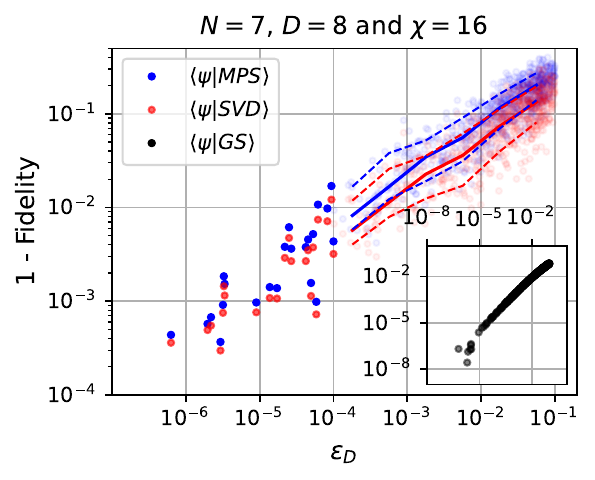}}}
\hfill
\subcaptionbox{}{\raisebox{+.09\height}{\includegraphics[width=0.31\textwidth,trim={0.26cm 0.27cm 0.3cm 0.25cm},clip]{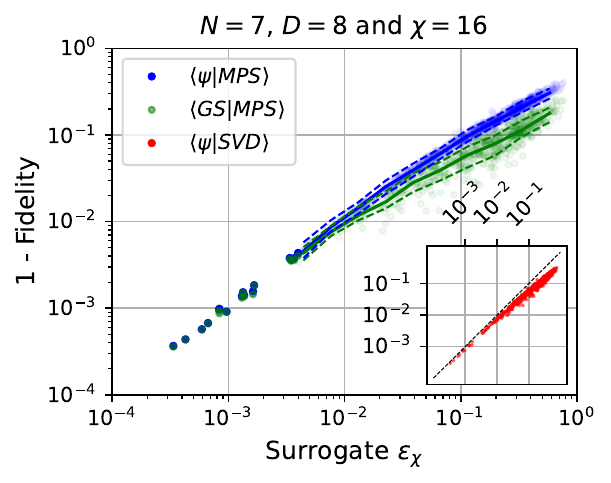}}}
\hfill
\subcaptionbox{}{\includegraphics[width=0.36\textwidth,trim={0.26cm 0.3cm 0.3cm 0.25cm},clip]{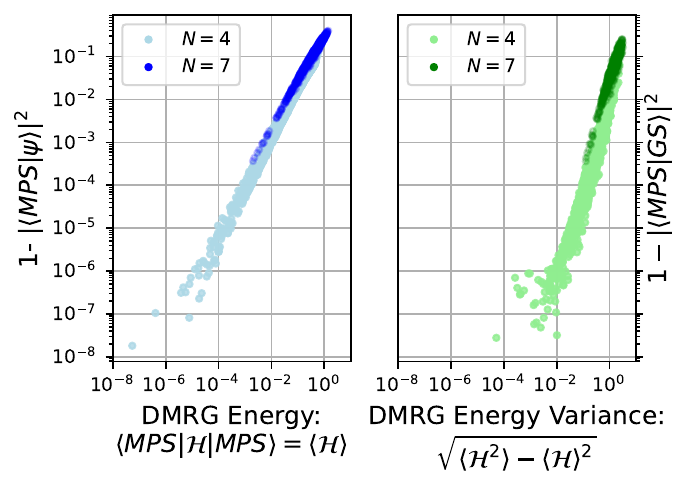}}
\caption{\label{fig:GBSErrors}Each point is a GBS simulation of a random pure state covariance matrix in the optimal local basis and with zero means. \textbf{(a)}\&\textbf{(b)} Due to the unbounded Hilbert space, "Fidelity" in this case refers to $\langle \alpha|\beta\rangle := \frac{|\langle \alpha|P|\beta\rangle|^2}{\langle P\rangle_\alpha\langle P\rangle_\beta}$, i.e. how much the states align in the truncated space. The full line is the mean of a bin of points while the dashed are the $\pm 31.4\%$ percentile (corresponding to $\pm$ one standard deviation for Gaussian distributions).
 "Surrogate $\varepsilon_\chi$" is the upper bound on $\varepsilon_\chi$, eq. \ref{eq:SurrogateEpsChi}. 
The fact that the normalized overlap between $|\psi\rangle$ and $|SVD\rangle$ is well correlated and bounded by this metric is an indicator that this is in fact a good surrogate for $\varepsilon_\chi$.
\textbf{(c)} In all cases $D=8$ and $\chi=16$. Clearly, the simulation errors can be estimated by the different moments of the variational Hamiltonian as explained in sec. \ref{sec:SimulationError}.
The asymptotic scaling of the data can be found in the SM sec. 7 \cite{supp}
}
\end{figure*}
The problem of Gaussian Boson sampling will now be treated with our variational approach. The problem consists of finding the ground state of eq. \ref{eq:GBSHamiltonian} in the optimal local basis described in sec. \ref{subsec:LBO}. The variational optimization algorithm is chosen to be single site DMRG as implemented in the python library Quimb \cite{gray2018quimb}.

GBS is well suited as a benchmarking problem, since the pure state amplitudes can be calculated by other means and compared against. For small system sizes (up to $D^N\sim 10^6$) the python library "thewalrus" is utilized \cite{Gupt2019}. For larger system sizes, one can compare to state of the art classical simulations of GBS, also based on Tensor Network methods, such as \cite{Oh2024}.

As noted in sec. \ref{sec:SimulationError}, there exists two simulation errors, originating from the cutoffs in the two hyper-parameters $D$ (physical dimension) and $\chi$ (bond dimension). In order to validate this statement and more importantly, the statement of eq. \ref{eq:MagOfSimError}, namely that all simulation errors are captured by the variational energy, we show the state overlaps as a function of the associated error amplitudes in figure \ref{fig:GBSErrors}. A schematic overview of the relevant states is found in eq. \ref{SchematicStates} and for the numerics, each state is to be understood as a normalized version of its projection wrt. $P$ (eq. \ref{eq:ProjectionOperator}). Therefore, the y-axis of figure \ref{fig:GBSErrors} shows to what degree the finite dimensional vectors align with each other.

The simulation error (colored blue in fig. \ref{fig:GBSErrors}), is directly correlated with the variational energy and how much of this is due to the finite bond dimension can be estimated by the variance of the variational energy (fig. \ref{fig:GBSErrors}c).
How $\varepsilon_D$ and $\varepsilon_\chi$ individually contribute to the simulation error is shown in fig. \ref{fig:GBSErrors}a and \ref{fig:GBSErrors}b with the insets establishing the correctness of these error metrics. The variationally obtained state is generally worse than the compressed state in simulating $|\psi\rangle$ due to finite a $\varepsilon_D$. This discrepancy can be mitigated by working towards approximating the effective Hamiltonian, eq. \ref{eq:EffectiveHamiltonian}.

\begin{figure}
    \centering
    \includegraphics[width=1\linewidth]{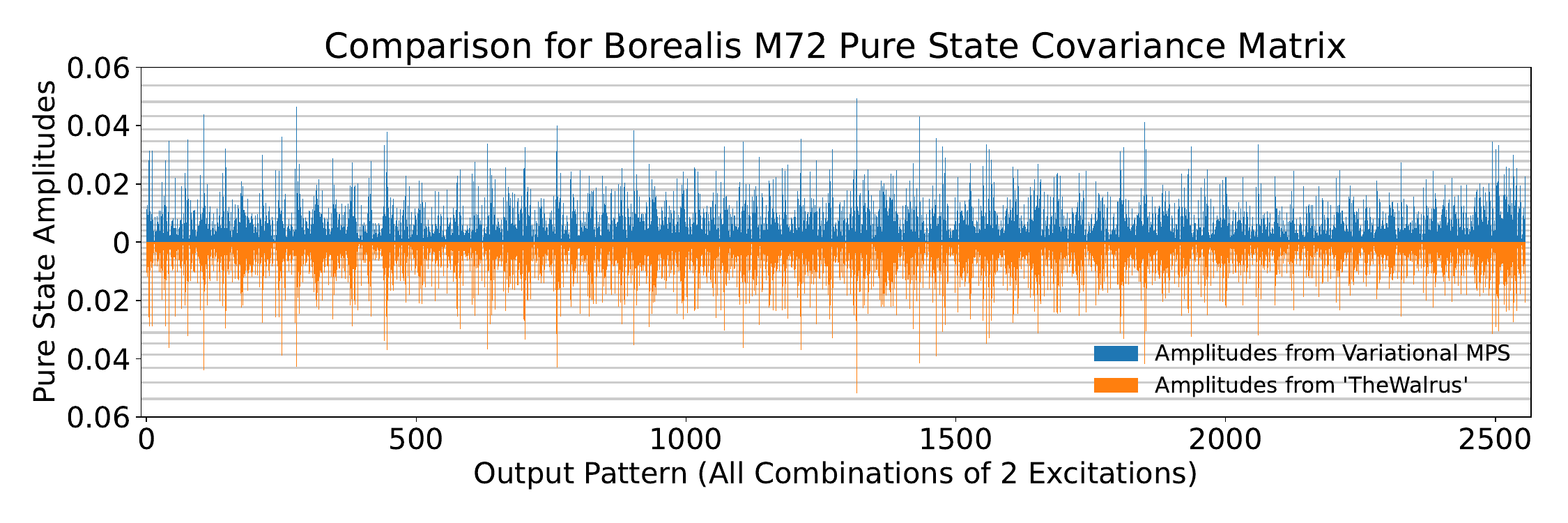}
    \caption{\label{fig:BorealisM72Comparison}
    Direct comparison of pure state amplitudes obtained through our variational approach to the true distribution for the Borealis M72 experiment \cite{Madsen2022-nq}, as defined by the $Q$ in eq. \ref{eq:CovarianceDecomposition}. The variational energy in this case is $\langle \mathcal{H} \rangle \approx 0.159$ while the normalized overlap with the compressed true state is: $|\langle SVD|MPS\rangle|^2 \approx 0.977$}
\end{figure}

We also compare our method to experiment, namely the Borealis M72 experiment \cite{Madsen2022-nq}.
This is done by using the procedure, described in sec. \ref{sec:SamplingWithNoise}, to obtain a pure state covariance matrix from the noisy experiment. Using our method with $D=6$ and $\chi=240$, we obtain the amplitude for all overlaps between the variational state and the associated Fock states. By focusing on only the two-excitation Fock state amplitudes, we can calculate the true target distribution analytically, as implemented in e.g. thewalrus. The comparison shows agreement as seen in fig. \ref{fig:BorealisM72Comparison}.

In order to investigate the benefit of the optimal local basis for GBS problems, we consider the total photon number distribution of the Borealis experiments in three different cases, as shown in table \ref{table:BorealisPhotonNumberDistributions}. The first column contains the collective photon number statistics for the pure states obtained from Williamson Decomposition of the mixed state covariance matrices of the noisy experiments \cite{PRXQuantum.3.010306}. The second column extracts the pure state with minimal expected number of excitations, through the use of Semi-Definite Programming (SDP) \cite{Oh2024}. This results in only a slight improvement when applying LBO (sec. \ref{subsec:LBO}) on the SDP-recovered pure state as shown in the last column.
\begin{table}
\caption{Photon Number Expectations \cite{PhysRevA.99.023817}}
$$\left(\sum_j \langle \hat{n}_j\rangle \pm \sqrt{\sum_j \langle \hat{n}_j^2\rangle - \langle \hat{n}_j\rangle^2}\right)$$
\begin{tabular}{l|l|l|l}
                 & Williamson                       & Semi-Definite             &Local Basis         \\
Experiment       & Decomposition                    & Programming               &Optimization        \\ \hline
BorealisM16      & $2.016 \pm 1.614 $               & $0.5487 \pm 0.807$        & $0.470 \pm 0.696$    \\
BorealisM72      & $7.072 \pm 2.895$                & $1.749 \pm 1.388$         & $1.601 \pm 1.280$    \\
BorealisM216High & $36.569 \pm 6.759$               & $6.541 \pm 2.686$         & $6.094 \pm 2.504$    \\
BorealisM288     & $46.930 \pm 7.585$               & $10.687 \pm 3.422$        & $10.106 \pm 3.235$                
\end{tabular}
\caption*{\label{table:BorealisPhotonNumberDistributions}The total photon number distribution for different pure states related to the experiments of the first column. The smaller the numbers, the easier it is to represent the state.
}
\end{table}

\subsection{Beyond GBS with parameterized LBO}
\begin{figure*}
\centering
\subcaptionbox{}{\includegraphics[width=0.31\textwidth,trim={0.26cm 0.28cm 0.3cm 0.25cm},clip]{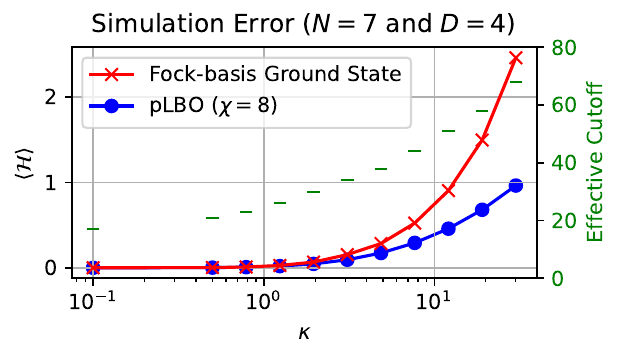}}
\hfill
\subcaptionbox{}{\includegraphics[width=0.31\textwidth,trim={0.26cm 0.27cm 0.3cm 0.25cm},clip]{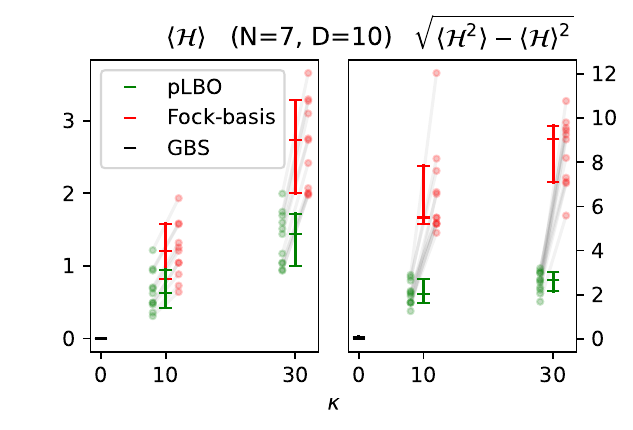}}
\hfill
\subcaptionbox{}{\includegraphics[width=0.36\textwidth,trim={0.26cm 0.3cm 0.3cm 0.25cm},clip]{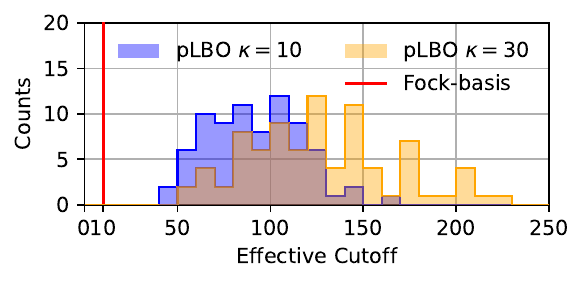}}
\caption{\label{fig:pLBOSimulations}
\textbf{(a)} The variational energy as a function of the "strength" ($\kappa$) of the non-Gaussian circuit for the Fock basis and the pLBO approach. The adaptability of the parameterized local basis increases the range of applicability of our variational approach for sampling problems.
Increasing $\chi$ in this case does not lower $\langle \mathcal{H}\rangle$, however, it still lowers $\sigma_\mathcal{H}$, meaning $\varepsilon_D$ is the dominant error.
\textbf{(b)} Simulation of eleven GBS problems with the non-Gaussian global CZ-gate applied, for increasing $\kappa$. The horizontal lines indicate the median and the $50\% \pm 34.1\%$th quantile. The grey lines group points that share the same GBS covariance matrix. With this we clearly see that pLBO always betters the simulation and increasingly so for more complex simulations.
\textbf{(c)} Distribution of the effective cutoffs (eq. \ref{eq:Effectivecutoff}) for the pLBO simulations of figure b. Recall that all computations for pLBO is done at a numerical cutoff of $D=10$, however, an increase of an order of magnitude is achieved in the corresponding cutoff in the Fock basis. Furthermore, the increase is not simply due to a large displacement or squeezing, but a learned consortium of the basis parameters, as we show in the supplemental material figure 9 \cite{supp}
}
\end{figure*}
Having established the validity of our variational method and usage of the variational energy as a measure of simulation error, we will now investigate the applicability of our pLBO approach (described in section \ref{subsec:pLBO}) by applying it to the non-Gaussian sampling problem defined in sec. \ref{subsec:VarHamilBeyondGBS}. 

The parametric local basis transformations are implemented directly in the matrix representation of eq. \ref{eq:SymplecticBasisTransformation}.
Consequently, we would like direct access to the basis parameters in the Hamiltonian, so that they can be updated at each iteration of the optimization algorithm. The most efficient way of doing so, is to have an algebraic construction of the MPO so that the MPO cores can quickly be generated with the new basis-parameters. The problem of writing down an MPO can be solved with Finite State Machines \cite{TimeEvolutionMethodsMPO}\cite{PhysRevA.78.012356}\cite{10.21468/SciPostPhys.3.5.035}. However, in the SM sec. 8 \cite{supp}, we present our manual approach for constructing the relevant MPOs for the non-Gaussian Hamiltonian eq. \ref{eq:nonGBSHamiltonian}

As a simple initial example, we consider the fully connected CZ-gate acting on the vacuum, such that the covariance matrix is the identity matrix and the displacements are zero. In this case, the only parameter defining the sampling problem is $\kappa$, with $\kappa=0$ being the trivial problem of sampling the vacuum. When $\kappa$ is raised, it will be increasingly more difficult to simulate the sampling problem, as captured by deviations in the variational energy $\langle \mathcal{H} \rangle$ from zero (sec. \ref{sec:SimulationError}). 
In figure \ref{fig:pLBOSimulations}a we show the variational energy for the pLBO approach compared to doing no basis optimization. The pLBO improves the simulation because it is able to diminish $\varepsilon_D$ by achieving a high effective cutoff.
$\varepsilon_\chi$ is practically negligible in this example since increasing $\chi$ had no effect on the variational energy. However, the Fock-basis energy is for the dense ground state ($|GS\rangle$).

If we apply the global CZ-gate on GBS problems instead, we achieve the results of figures \ref{fig:pLBOSimulations}b and \ref{fig:pLBOSimulations}c.
Each point is a randomly generated pure state covariance matrix, post-selected to not deviate too far from the identity matrix, which results in the GBS problem being easily simulated. However, raising $\kappa$ increases the complexity. Still, pLBO is better able to simulate the sampling problem with minimal overhead. The reason is again the increased cutoff as seen in fig. \ref{fig:pLBOSimulations}c.
Interestingly the variance of the Hamiltonian and thus $\varepsilon_\chi'$ is affected by the application of the single site gates of pLBO. While single site gates cannot change the entanglement structure, they will affect the truncation error $\varepsilon_D$ which in turn affect the ground state $|GS(\varepsilon_D)\rangle$ and its entanglement structure.

\begin{figure}
    \centering
    \includegraphics[width=\linewidth]{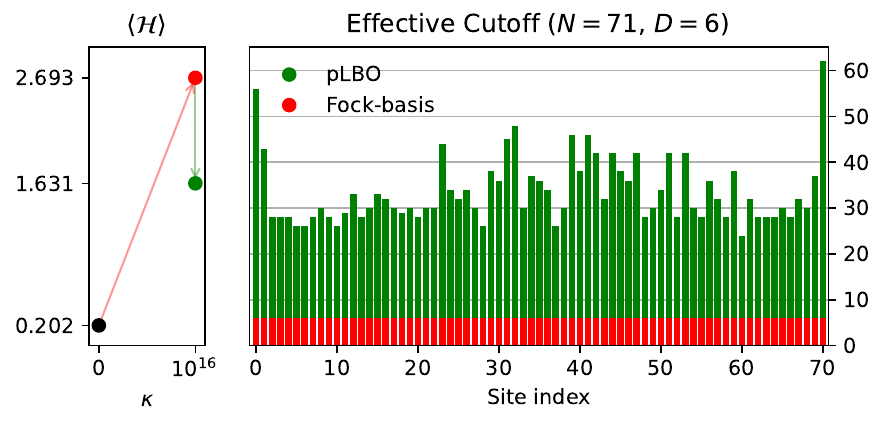}
    \caption{\label{fig:Comparison_pLBO_GS_N71}Application of our pLBO approach to a sampling problem with $N=71$ modes, showcasing its scalability. For GBS ($\kappa = 0$) $\chi=240$ whereas for the non Gaussian case the optimization converged already at $\chi=64$. The reason for the exponentially large $\kappa$ in this case is due to the parameterization of the global CZ-gate not accounting for the number of modes: $\exp(-i\kappa\prod_n x_n) = \exp(-i\prod_n\sqrt[N]{\kappa}x_n)$.}
\end{figure}

To prove the scaleability of our approach, we apply our method to a likewise sampling problem with $N=71$, as shown in fig. \ref{fig:Comparison_pLBO_GS_N71}. 
Similar results are found in this moderately large $N$ regime and indicates the usefulness of LBO techniques for bosonic variational problems.

\section{\label{sec:discussion}Discussion}

\subsubsection{Gaussian Boson Sampling}
State of the art simulation of GBS \cite{Oh2024} compresses the Gaussian state to MPS form through the use of sequential singular value decompositions \cite{PhysRevLett.91.147902}. 
While such a single-shot compression technique is quasi-optimal (see SM sec. 3), it is generally the case, that the MPS obtained from iterative methods, such as DMRG, will better approximate the optimal MPS representation \cite{Khoromskij+2018}\footnote{While precise studies on this topic has eluded our attention, an accessible prototypical example is given in \cite{T3F}, where the optimization algorithm produces a better compression than the single-shot SVD-based compression.}.
Therefore, if the error $\varepsilon_D$ is eliminated, e.g. by deriving the exact effective Hamiltonian (eq. \ref{eq:EffectiveHamiltonian}), our method provides a way to obtain the optimal MPS representation of general continuous variable sampling problems.
While efforts towards obtaining the effective Hamiltonian could help alleviate the $\varepsilon_D$ error another approach consists of utilizing the symmetries present in the sampling problems \cite{PhysRevA.108.052604}, thereby effectively increasing the number of basis states \cite{Xiang_2023}.

\subsubsection{Optimization strategies and gradient based optimization algorithms}
By framing the sampling problem as a variational problem we can leverage all the techniques developed for solving such optimization problems.
While we will not provide an extensive overview of such methods, we will here touch on our experiences, especially wrt. the parameterized Local Basis Optimization.

Since we are unable to employ linear methods (due to eq. \ref{eq:SymplecticBasisTransformation}), we opt for gradient based optimization methods, where the gradients are obtained by auto differentiation.
If we were to optimize the whole state at once (akin to what is done in \cite{RAKHUBA2019718}), the memory consumption of the gradient calculation would become the limiting factor. Instead, each site is optimized sequentially akin to single site DMRG. Seemingly, this has a regularizing effect and yields consistently equal or better minima than optimizing all variational parameters at once. However, further studies would have to distinguish whether this is dependent on the specific optimization algorithm.

As mentioned earlier, we have omitted the $(D\times d)$ linear transformation which would possibly diminish the critical error $\varepsilon_D$. We do this to focus the study on the variational parameters that are implemented algebraically. Still, in the supplemental material \cite{supp} sec. 6 we apply a single round of linear basis transformation optimization to the results of fig. \ref{fig:Comparison_pLBO_GS_N71}. The gain is larger for the Fock-basis simulation since the pLBO already has enhanced the basis. However, we expect that with a simultaneous optimization of the two methods, they will be better suited to leverage each other and an even higher effective cutoff may be reached. This remains for future studies to verify.

Regarding optimization algorithms, the division of the optimization into sub-problems is the setting for Stochastic Gradient Descent. Within this group of algorithms, ADAM \cite{kingma2017adammethodstochasticoptimization} is a popular choice and we also found this to perform better than non-linear Conjugate Methods in reaching better minima, but at the cost of more iterations.
L-BFGS \cite{Liu1989} was also found to converge fast, but only to local minima, from which SGD could be employed to converge towards a better minimum.
We believe this "initially fast then slow" behavior of L-BFGS is related to the considerations necessary when implementing DMRG \cite{TensorNetworkOrg}. Namely, that the immediate site should not be fully optimized within a sweep, since the effective Hamiltonian is only tentative as the other sites have yet to converge. This is in conflict with the line-search employed during a step of L-BFGS. 

Using DMRG, the bond dimension $\chi$ is iteratively increased from $\chi = \mathcal{O}(1)$ to the wanted final bond dimension \cite{PhysRevB.91.155115}. Doing this increases the speed of convergence, presumably due to a smaller need for large Krylov spaces, and it could be interesting to also implement the subspace expansion method in our optimization procedure \cite{doi:10.1137/140953289}\cite{10.1007/978-3-319-10705-9_33}. However, at least for the present non-Gaussian sampling problem, we found that running pLBO at a small bond dimension ($\chi = \mathcal{O}(D)$) yielded equally good local basi as running it for the maximum bond dimension. 
Thus, we propose the following optimization procedure that best incorporates the strengths of both DMRG and the pLBO approach:
\begin{itemize}
    \item Use DMRG in the Fock basis to produce a good initial guess for pLBO by sweeping until $\chi=\mathcal{O}(1) \rightarrow \chi = \mathcal{O}(D)$
    \item Optimize the local basis together with the variational state using Stochastic Gradient Descent.
    \item Use DMRG in the obtained optimal local basis to converge towards the optimal variational state with the maximal achievable bond dimension.
\end{itemize}
In fact, all results for the non-Gaussian example, except those shown in fig. \ref{fig:pLBOSimulations}a, were produced with this procedure.

\subsubsection{Further direction of studies for pLBO}
The parameterization of the local basis transformation eq. \ref{eq:pLBOUnitary} is only one choice of many. The only overhead exists in the implementation of the matrix representation (see SM sec. 5 \cite{supp}). Therefore, it could be worthwhile to better understand the variational class spanned by the different parameterizations, in order to pick the optimal one.

In order to truly leverage the non-linear capabilities, we will now speculate in methods that would be able to also reduce the computational bond dimension, akin to how the pLBO is able to reduce the computational physical dimension. One way of achieving this is to consider a parameterized classically simulateable circuit, e.g. a low-depth application of two-site gates. The transformation of the Hamiltonian under such a circuit would produce a collection of new terms with the coefficients depending non-linearly on the introduced parameters. Using Finite State Machines, one would be able to formulate the MPO in such a way, that the tuneable parameters are easily accessed and can thus be iteratively optimized using gradient based methods. The result is a ground state optimization algorithm that simultaneously learns an MPS representation, but which necessitates a smaller bond- physical-dimension than in the original formulation, together with the parameters for a small depth circuit that can be applied to retrieve the state in the original representation. Similar ideas of divide-and-conquer are utilized in simulating qubit-based quantum circuits in the form of a joint Schrödinger-Heisenberg picture approach \cite{KENNES201637}\cite{doi:10.1126/sciadv.adk4321}\cite{PRXQuantum.5.010308}.
Lastly, we emphasize that our simulation tool can be straightforwardly applied to the setup in \cite{PhysRevA.106.042413},
which may help determine the noise threshold for efficient classical approximate simulations in non-linear Boson Sampling.

\begin{acknowledgments}
The authors would like to thank Rafael E. Barfknecht for his contributions in initializing the project. We further thank  Andreas Michelsen, Jonas Neergaard-Nielsen and Karsten Flensberg for useful discussions. 
We acknowledge support from the Carlsberg foundation and from the Novo Nordisk Foundation, Grant number NNF22SA0081175, Quantum Computing
Programme.
\\

{\footnotesize \flushleft The data and the code used to generate the results are accessible on GitHub at: 
github.com/Vinther901/Variational-Tensor-Networks-Boson-Sampling \cite{githubrepo}}
\end{acknowledgments}

\renewcommand{\appendixname}{Supplemental Material}
\appendix*

\section{}
\subsection{
Effect of truncation on the spectrum of $\mathcal{H}$}
Equation \ref{eq:CentralEquation} and \ref{eq:generalizedCCR} results in: $\text{spec}(\mathcal{H}) \in \{0,1,2,\hdots\}=\mathbbm{N}_0$. Therefore, the distribution of spectral weight for the variational state $|MPS\rangle$ can be alluded to by:
\begin{equation}
    \langle \mathcal{H}\rangle_{|MPS\rangle} = 0\cdot |c_0|^2 + \sum_{m=1}^\infty m|c_m|^2
\end{equation}
where $\sum_{m=0}|c_m|^2=1$ and $|c_0|^2$ is the object of interest. The $|c_m|^2$ is the total amplitude for all degenerate eigenvectors corresponding to the eigenvalue $m$ and, in principle, the $|c_m|^2$'s could be fitted by also considering higher moments $\langle \mathcal{H}^n\rangle$.

\begin{figure}[ht]
    \centering
    \includegraphics[width=1\linewidth, trim={0cm 3cm 7.4cm 11cm},clip]{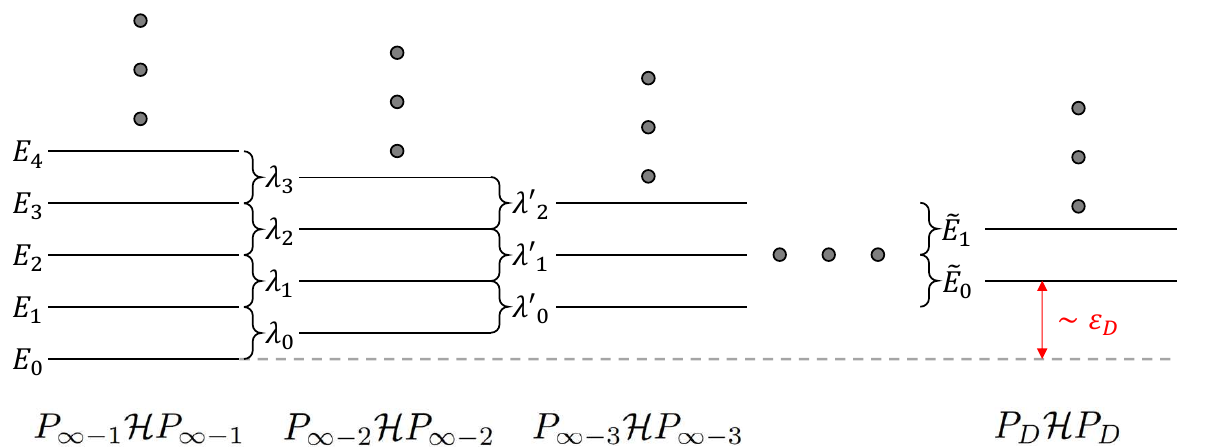}
    \caption{\label{appendix:figPerturbingTheEnergy}Outline of how the energies of the truncated Hamiltonian is related to the original spectrum. $P_{\infty-1}$ is a projection onto a large enough Hilbert space such that the relevant part of the spectrum remain unaffected. The result of Cauchy's interlace theorem \cite{fbb19167-de4e-311f-b329-ce5946170d56} \cite{Horn_Johnson_1985} is that $\lambda_0 \leq\lambda_0'\leq\lambda_1\leq\lambda_1'\leq\hdots$ as exemplified by the brackets in the figure. The vertical lines are just one arbitrary manifestation and should not be taken too seriously (except for the $E_i$'s which truly are harmonic). The value of $\tilde{E}_0$ is investigated in fig. \ref{appendix:figTildeE0}.
    }
\end{figure}

One could be concerned that once the projections are introduced and thus the spectrum altered, this will no longer be true. However, it can be shown (fig. \ref{appendix:figPerturbingTheEnergy}), that introducing the projections can only increase the eigenvalues \cite{Horn_Johnson_1985}. Furthermore, when they do increase, it is because the truncation $P$ introduced an error. Therefore, this metric remains trustworthy and a variational energy $\langle \mathcal{H}\rangle_{P|MPS\rangle}<1$ determines an acceptable simulation, depending on the wanted degree of accuracy. A larger variational energy could still result in an acceptable simulation, however, it would be harder to recognize.

\begin{figure}[ht]
    \centering
    \includegraphics[width=0.5\linewidth]{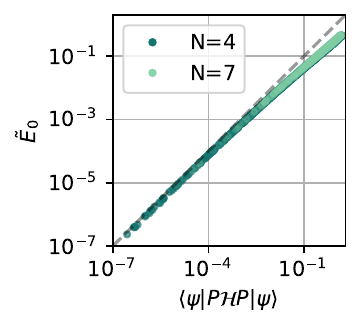}
    \caption{To first order in perturbation theory $\Tilde{E}_0 \lesssim \langle \psi |P\mathcal{H}P|\psi\rangle$ (with $E_0=0$ and the inequality from numerical observation).
Indeed, the ground state energy forms a very nice line as a function of $\langle\mathcal{H}\rangle_{P|\psi\rangle}$}
\label{appendix:figTildeE0}
\end{figure}

\subsection{
Continuous Variable construction of $\mathcal{H}$}
Suppose we want to sample from a few-body wavefunction of the form:
\begin{equation}
    \psi(\pmb{x}) \propto e^{-\varphi(\pmb{x})}
\end{equation}
We can then define a generalized ladder operator:
\begin{equation}
    \hat{A}_k =\sum_l M^{-1/2}_{kl}\left(\hat{P}_l - i\frac{\partial\varphi(\pmb{x})}{\partial x_l}\bigg|_{\pmb{x}=\hat{\pmb{X}}}\right)
\end{equation}
or in vector notation:
\begin{equation}
\label{appendix:eqGeneralizedLoweringOp}
    \pmb{\hat{A}} = M^{-1/2}(\hat{\pmb{P}} - i\nabla\varphi(\hat{\pmb{X}})),
\end{equation}
where $\hat{\pmb{X}},\hat{\pmb{P}}$ are the canonical position and momentum operators. 
The positive semi-definite matrix $M$ is the mass-tensor. It is complimentary to the wavefunction and not given from experiment.
The generalized ladder operators eq. \ref{appendix:eqGeneralizedLoweringOp} are related to those found in super-symmetric quantum mechanics \cite{Drigo_Filho_2001}

If we set the mass matrix to be the identity:
\begin{equation}
    \hat{A}_k = \hat{P}_k - i\frac{\partial\varphi(\pmb{x})}{\partial x_k}\bigg|_{\pmb{x}=\hat{\pmb{X}}}
\end{equation}
then it is still true that $\pmb{\hat{A}}|\psi\rangle = 0$ and that $\pmb{\hat{A}}^\dagger\pmb{\hat{A}}$ is positive semi-definite.
However, with the introduction of a cutoff: $\hat{P} = \sum_k^{D-1}|\pmb{k}\rangle\langle\pmb{k}|$ we generally have, $\hat{P}\hat{A}_n\hat{P}|\psi\rangle = |\varphi_n\rangle$ where $|\varphi_n\rangle=0$, only when no truncation error is made from the projection.
Notice we still have positive semi-definiteness:
\begin{equation}
    \langle \pmb{\hat{A}}^\dagger\pmb{\hat{A}}\rangle_{P\psi} = \sum_n \langle \varphi_n|\varphi_n\rangle
\end{equation}
with each $\langle \varphi_n|\varphi_n\rangle\geq 0$. We can then consider the introduction of an arbitrary symmetric and positive-definite matrix, denoted the mass matrix $M$: 
\begin{equation}
\langle \pmb{\hat{A}}^\dagger\pmb{\hat{A}}\rangle_{P\psi} \rightarrow \langle \pmb{\hat{A}}^\dagger M^{-1}\pmb{\hat{A}}\rangle_{P\psi}
\end{equation}
such that
\begin{equation}
    \langle \pmb{\hat{A}}^\dagger M^{-1}\pmb{\hat{A}}\rangle_{P\psi} = 
    \sum_{nml} \langle \varphi_n|M^{-1/2}_{nl}M^{-1/2}_{lm}|\varphi_m\rangle =
    \sum_l \langle \Tilde{\varphi}_l|\Tilde{\varphi}_l\rangle
\end{equation}
which means that positive semi-definiteness is still obeyed but also that, once a truncation $P$ is introduced, the simulation can be affected by the choice of $M$. Furthermore, the choice of $M$ also has an effect on the commutation relations:
\begin{align}
    [\hat{A}_n,\hat{A}_m] &= 0\\
    [\hat{A}_n,\hat{A}^\dagger_m] &= 2M_{ni}^{-1/2}\text{Re}\left(\frac{\partial^2\varphi(\hat{\pmb{X}})}{\partial x_i\partial x_j}\right)M_{jm}^{-1/2}
\end{align}
such that if $\text{Re}(\varphi(\pmb{x}))$ is at most quadratic, $M$ can be chosen such that $[\hat{\pmb{A}},\hat{\pmb{A}}^\dagger] = \mathbbm{1}$. In the case of GBS and with this choice of $M$, a construction of $\mathcal{H}$ from eq. \ref{appendix:eqGeneralizedLoweringOp} yields exactly eq. \ref{eq:GBSHamiltonian}.

Empirically (at least for GBS) the $M$ that make the generalized ladder operators obey the canonical commutation relations also improves the simulation, as opposed to choosing $M$ to be the identity.
In general however, numerics indicate that even though it provides an improvement, it will not be the optimal choice. The optimal choice remains elusive and it also seems to dependent on the specific choice of $D$.

\subsection{\label{appendix:MPSCompression}Compression to MPS form}
Imposing an MPS form on a state, with maximum bond dimension $\chi$, usually incurs an approximation error.
In order to define this error, first note that MPSs of a given bond dimension constitute a smooth manifold \cite{10.1063/1.4862851}
\begin{equation}
    \mathcal{M}_\chi = \{|\Phi\rangle : \text{MPS-rank}(|\Phi\rangle) = \chi\}
\end{equation}
$\text{MPS-rank}(|\Phi\rangle) = \chi$ means that $|\Phi\rangle$ can be decomposed as eq. \ref{eq:GenericState} and \ref{eq:MPSAnsatz} with the predefined bond dimension structure $\chi$.
By compression of a generic state $|\psi\rangle$ to its MPS form $|\psi_\text{MPS}\rangle$, we mean a retraction from the point generally outside the manifold onto the manifold \cite{SCHOLLWOCK201196}.
\begin{equation}
    |\psi_\text{MPS}\rangle = \underset{|\Phi\rangle \in \mathcal{M}_\chi}{\mathrm{argmin}}\big|\big||\psi\rangle - |\Phi\rangle\big|\big|^2
\end{equation}
There exist various approximations to $|\psi_\text{MPS}\rangle$ \cite{SCHOLLWOCK201196}\cite{PhysRevB.110.085149}. The one used in this paper relies on sequential singular value decompositions \cite{PhysRevLett.91.147902}\cite{Oh2024}. 
\begin{equation}
\label{eq:SVDApproximation}
   |\psi_\text{MPS}\rangle \approx \Tilde{|SVD\rangle}
   ,\quad
   |SVD\rangle = \frac{\Tilde{|SVD\rangle}}{\sqrt{\Tilde{\langle SVD}\Tilde{|SVD\rangle}}}
\end{equation}
hence the name "SVD" for the compressed state. The tilde is to indicate that the state generally will not be normalized and a subsequent renormalization is necessary.
Diagrammatically, if we denote a singular value decomposition by a dashed line, we may write:
\begin{equation}
\label{eq:SVDApproximationDiagram}
   \adjincludegraphics[valign=c,width=0.4\linewidth]{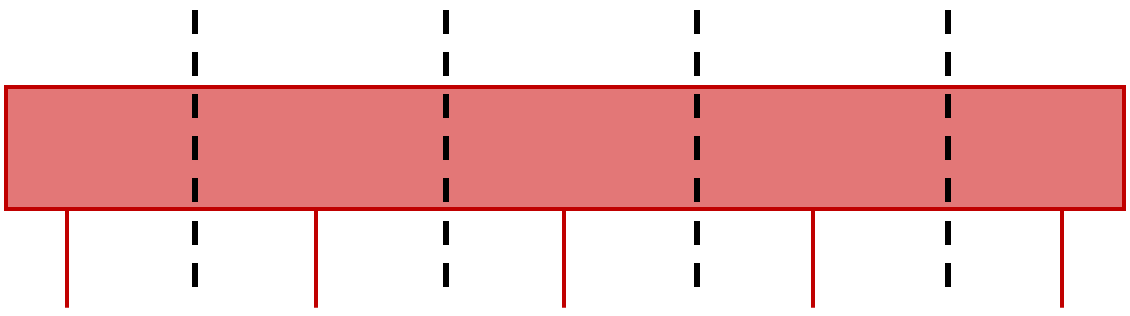} \approx \adjincludegraphics[valign=c,width=0.4\linewidth]{MPS_figs/MPSN5.png}
\end{equation}
where the approximation is due to discarding all but the $\chi$ largest singular values. Note that the SVDs cannot be carried out simultaneously and it is this arbitrariness that demotes it from being an optimal rank $\chi$ compression.

Similarly to $\varepsilon_D$ we define:
\begin{equation}
\label{eq:epsilonchi}
   \varepsilon_\chi := \big|\big| |\psi\rangle - \Tilde{|SVD\rangle} \big|\big|^2
\end{equation}
which obeys the following bound \cite{OSELEDETS201070}\cite{doi:10.1137/090752286}\cite{doi:10.1137/S0895479896305696}:
\begin{equation}
\label{eq:QuasiOptimality}
    \varepsilon_\chi \leq \sum_{n=1}^{N-1} \sum_{k> \chi_i} \left(\sigma ^{(n)}_k\right)^2  \leq (N-1)\big|\big| |\psi\rangle - |\psi_\text{MPS}\rangle \big|\big|^2
\end{equation}
where $\sigma^{(n)}$ are the Schmidt coefficients when bipartitioning the state into $\{1,\hdots,n\}$ and $\{n+1,\hdots,N\}$. The fact that the compression by sequential SVDs is only quasi-optimal, is made manifest by the second inequality.
The first inequality can also be found in the physics literature \cite{PhysRevB.73.094423}\cite{PhysRevB.110.085149}\cite{RevModPhys.93.045003}\cite{PhysRevX.10.041038}.
In practice, we circumvent the complications of the infinite dimensional Hilbert space in the definition of $\varepsilon_\chi$, by using the sum of squared singular values as a surrogate, eq. \ref{eq:SurrogateEpsChi}.

\subsection{\label{appendix:secTruncationErrorBound}Proof of $\varepsilon_D$ bound (eq. \ref{eq:SpecWfromSDsums})}
Similar to eq. \ref{eq:QuasiOptimality}, the subsequent application of SVDs in the LBO procedure has its own quasi-optimality condition associated to it:
\begin{equation}
\label{eq:QuasiOptimality2}
    \big|\big||\psi\rangle - |\psi_\text{LBO}\rangle\big|\big|^2 \leq \sum_{i=1}^N \sum_{k> d_i} \left(\sigma ^{(i)}_k\right)^2 \leq N \big|\big||\psi\rangle - |\psi_\text{opt}\rangle\big|\big|^2
\end{equation}
where, now, $\sigma ^{(i)}_k$ are the eigenvalues of eq. \ref{eq:ReducedDensityMatrix}, $|\psi_\text{LBO}\rangle$ is the physical state after truncation of the physical dimension in the optimal local basis and $|\psi_\text{opt}\rangle$ is the associated best rank $\{r_i\}$ approximation of the state $|\psi\rangle$.

However, the LBO based on the diagonalization of the reduced density matrices of the exact target state, is the optimal one \cite{PhysRevB.62.R747}, which means that:
\begin{equation}
\label{appendix:eqSVDOptimality}
    |\psi_\text{LBO}\rangle = P|\psi\rangle = |\psi_\text{opt}\rangle
\end{equation}
\\and the quasi-optimality condition eq. \ref{eq:QuasiOptimality2} becomes eq. \ref{eq:SpecWfromSDsums} by the two observations:
\begin{align}
    |||\psi\rangle - P|\psi\rangle||^2
    &= 1 - \langle \psi | P |\psi \rangle = \varepsilon_D \\
    \quad \sum_{k=0}^\infty \left(\sigma_k^{(i)}\right)^2 &= 1
\end{align}
One way of seeing \ref{appendix:eqSVDOptimality} is to note that what made eq. \ref{eq:SVDApproximation} and eq. \ref{eq:SVDApproximationDiagram} approximate, is that each sequential SVD is affected by the one before it. When considering eq. \ref{eq:SVDPhysicalDimension} this is no longer the case, each SVD can be performed in unison and the compression will be the optimal one \cite{PhysRevB.62.R747}.

\subsection{\label{appendix:secImplementingTransformedOperators}Numerical implementation of the transformed bosonic operators eq. \ref{eq:SymplecticBasisTransformation}}
The inability of a matrix representation of $\hat{X}$ and $\hat{P}$ to obey the canonical commutation relation is easily seen by taking the trace:
\begin{equation}
    \text{Tr}([X,P]) = 0 \neq i\hbar\text{Tr}(\mathbbm{1})
\end{equation}
In the SQHO basis the "error" will accumulate near the point of truncation:
\begin{equation}
    [X,P] \doteq i\hbar
    \begin{pmatrix}
        1 & 0 \\
        0 & \ddots & \ddots \\
        & \ddots &  1 & 0 \\
        & & 0 & 1-D \\
    \end{pmatrix}
\end{equation}
Therefore, if we rely on matrix multiplication to construct the matrix representation of eq. \ref{eq:SymplecticBasisTransformation} we should be wary of this defect.

From eq. \ref{eq:SymplecticBasisTransformation} we see that the sub/super-diagonal matrices $a$ and $a^\dagger$ only appears quadratically. The Hamiltonian eq. \ref{appendix:eqNonGaussianHamiltonian} is also only quadratic in $X$ or $P$ such that $a$ and $a^\dagger$ only appears to the fourth power in any computation. Therefore, we can implement a matrix representation of eq. \ref{eq:SymplecticBasisTransformation} simply by initializing a matrix representation of dimension $D+4$, construct the necessary Hamiltonian terms of the MPO through matrix multiplication and then truncating the matrix representations from $D+4\rightarrow D$, thereby discarding the faulty entries.

\begin{figure}
    \centering
    \includegraphics[width=\linewidth]{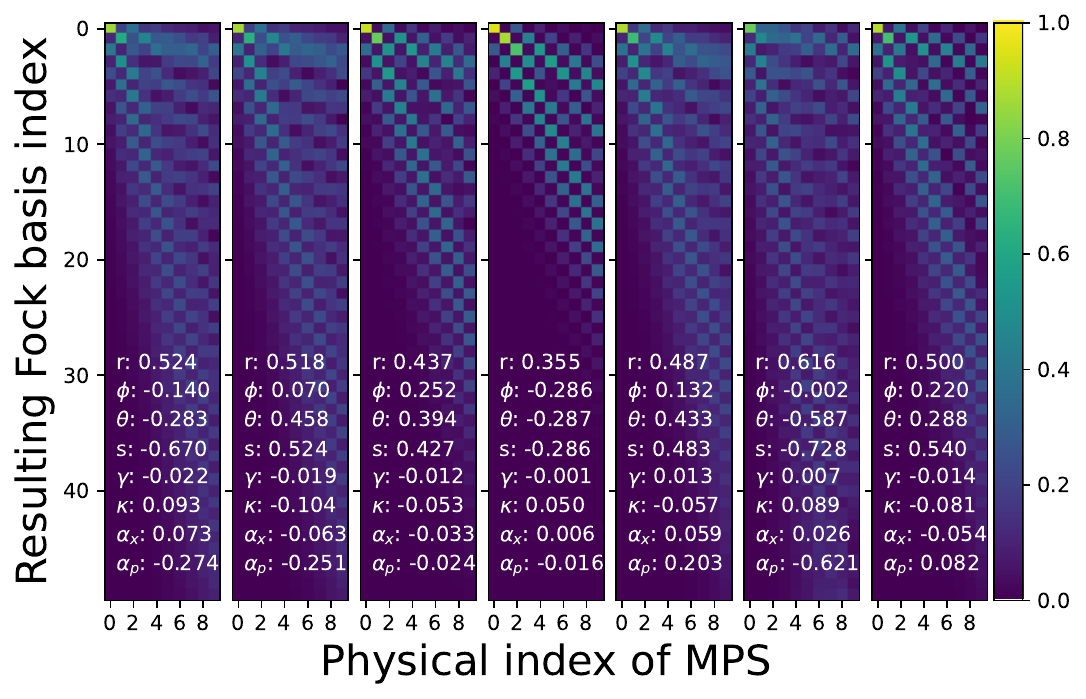}
    \caption{\label{appendix:figExampleBasisChange}Absolute value of the entries of the matrices $U_{nm}^{(i)}$ used to transform the variational state from the optimized local basis $|m_i\rangle$ to the Fock basis $|n_i\rangle$: $U_{nm}^{(i)} = \langle n_i|\mathcal{U}^{(i)}|m_i\rangle$
    This example is taken from one of the results in fig. \ref{fig:pLBOSimulations} with a large effective cutoff.
    }
\end{figure}

\subsection{\label{appendix:secLBO}Unifying Parameterized- and Linear Basis Optimization}
Presented in figure \ref{appendix:eqLBO} is the result of fig. \ref{fig:Comparison_pLBO_GS_N71} with the simulations of the non Gaussian experiments subjected to a round of Local Basis Optimization similar to the approach of Ref. \cite{PhysRevLett.108.160401}.
Specifically, starting from the MPS with physical dimension $d$ (namely those that produced the results corresponding to the filled dots), the linear isometry:
\begin{equation}
\adjincludegraphics[valign=c,width=0.08\linewidth]{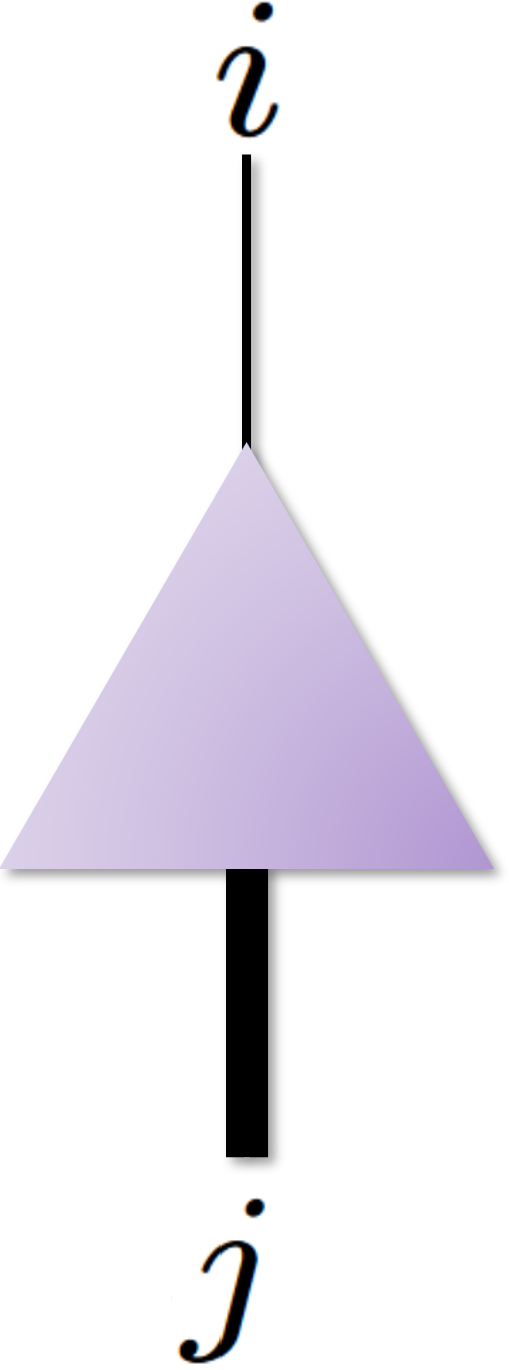} = 
    \begin{pmatrix}
        \mathbbm{1}_{d\times d} \\
        \pmb{0}_{(D-d)\times d}
    \end{pmatrix}_{ji}
\label{appendix:eqLBO}
\end{equation}
\begin{figure}
    \centering
    \includegraphics[width=0.25\linewidth]{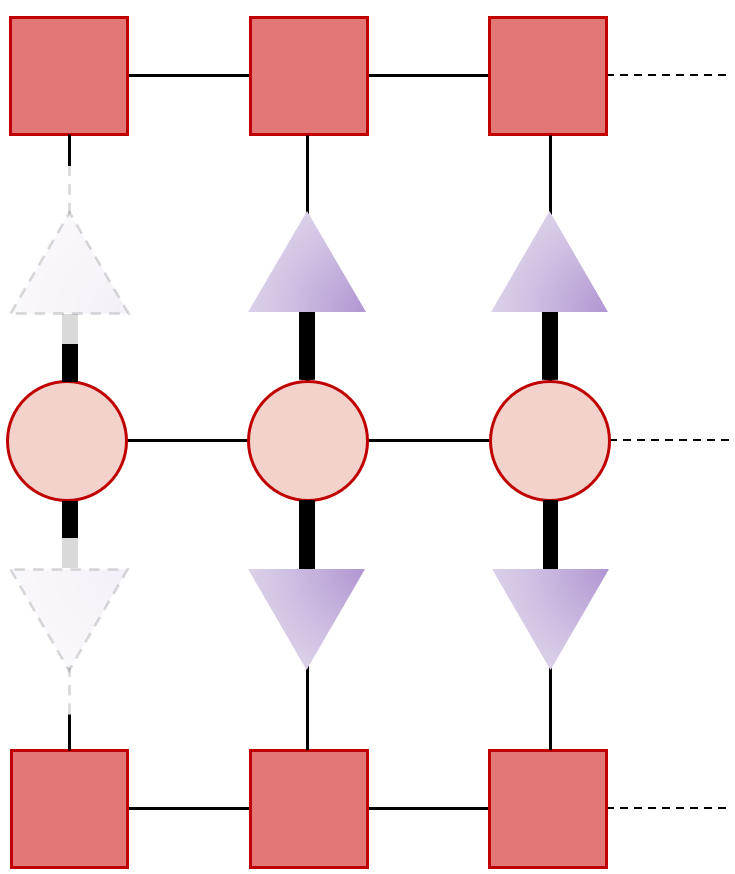}
    \caption{\centering Effective Hamiltonian for isometries.}
    \label{appendix:figEffHamilIsometry}
\end{figure}
\begin{figure}
    \centering
    \includegraphics[width=0.55\linewidth]{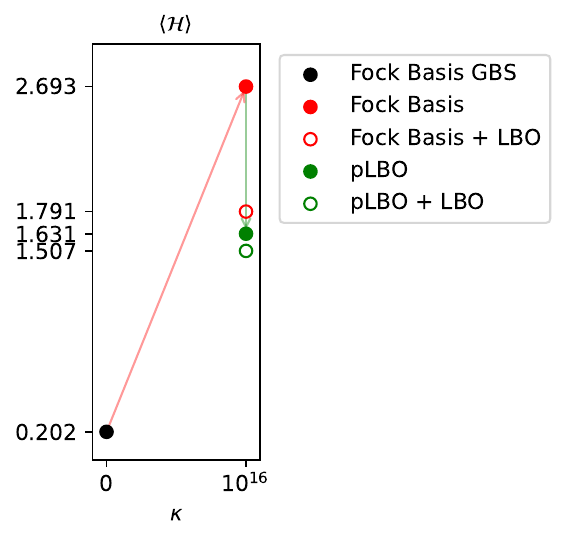}
    \caption{\label{appendix:figLBO}The results of this figure indicates how well the parameterized local basis optimization approach can account for any variance in the basis used for computation. The variational states that correspond to the filled dots have been subjected to a single sweep of standard local basis optimization and a subsequent DMRG optimization of the corresponding MPS in the optimized basis. The obtained and improved results, after such an optimization procedure, are shown by unfilled dots. As seen, the pLBO already accounts for a lot of the variation and the gain from LBO is smaller than for the Fock-basis.}
\end{figure}
is applied to each leg, effectively increasing the physical dimension from $d$ to $D$.
The isometries are optimized similar to how the MPS is optimized in DMRG. 
Starting from e.g. the left, an effective Hamiltonian is defined for the left-most isometry, as seen in fig. \ref{appendix:figEffHamilIsometry}.
To optimize it, is to replace it with the lowest energy state of the effective Hamiltonian. 
All the isometries are optimized in a sweeping manner and after obtaining the optimal isometries the MPS is optimized with DMRG in the new $d$ dimensional basis.
The unfilled dots of fig. \ref{appendix:figLBO} correspond to the results obtained after one such sweep of updating the isometries and running DMRG.
Note that this is not the same procedure as applied in \cite{PhysRevLett.108.160401}. Furthermore, the parameterized- and standard Local Basis Optimization doesn't have to be applied separately. In fact, we suspect the optimal method would be to use L-BFGS to learn the isometry (eq. \ref{appendix:eqLBO}) simultaneously with the variational basis parameters implemented in the MPO and then use linear methods for learning the matrices of the MPS. However, this remains as a question for further studies.

\begin{widetext}
\subsection{\label{appendix:secFig4AsymptoticScaling}Fig. 4 Asymptotic Scaling}
\begin{figure*}[ht]
\centering
\subcaptionbox{$
\begin{aligned}
    1 - F_{\langle \psi|MPS\rangle} &\sim 0.9029 \times \varepsilon_D^{0.5347}\\
    1 - F_{\langle \psi|SVD\rangle} &\sim 0.5496  \times \varepsilon_D^{0.5201}
\end{aligned}$
}{\raisebox{+.09\height}{\includegraphics[width=0.31\textwidth,trim={0.26cm 0.28cm 0.3cm 0.25cm},clip]{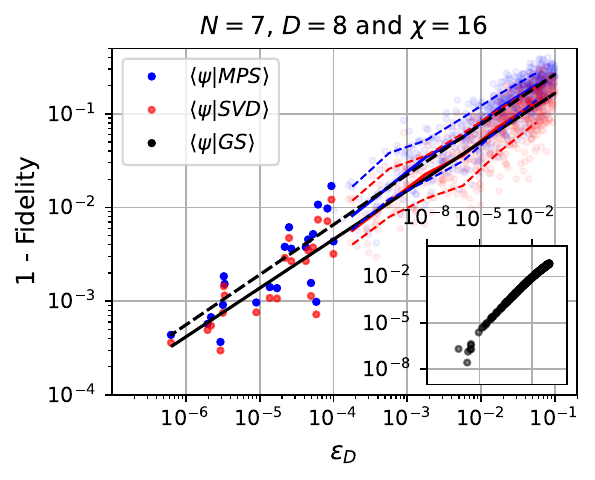}}}
\hfill
\subcaptionbox{$
\begin{aligned}
    &1 - F_{\langle \psi|MPS\rangle} \sim 0.5775 \times \varepsilon_\chi^{0.8563}\\
    &1 - F_{\langle GS|MPS\rangle} \sim 0.2721 \times \varepsilon_\chi^{0.7461}
\end{aligned}$
}{\raisebox{+.09\height}{\includegraphics[width=0.31\textwidth,trim={0.26cm 0.27cm 0.3cm 0.25cm},clip]{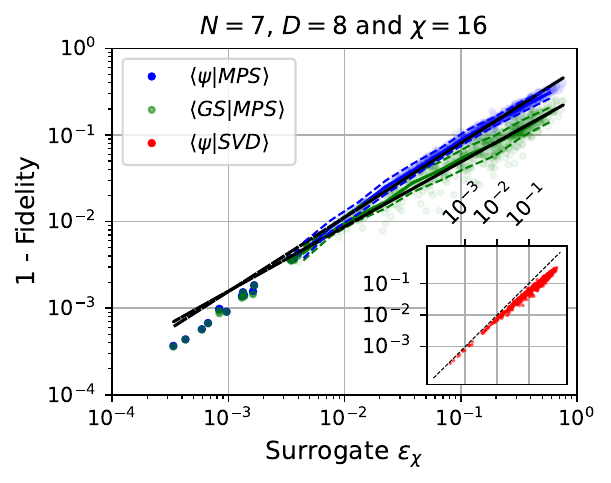}}}
\hfill
\subcaptionbox{$
\begin{aligned}
    &1 - F_{\langle \psi|MPS\rangle} \sim 0.2901 \times \langle\mathcal{H}\rangle^{1.1827}\\
    &1 - F_{\langle GS|MPS\rangle} \sim 0.0145 \times \sqrt{\langle\mathcal{H}^2\rangle-\langle\mathcal{H}\rangle^2}^{2.1833}
\end{aligned}$}{\includegraphics[width=0.36\textwidth,trim={0.26cm 0.3cm 0.3cm 0.25cm},clip]{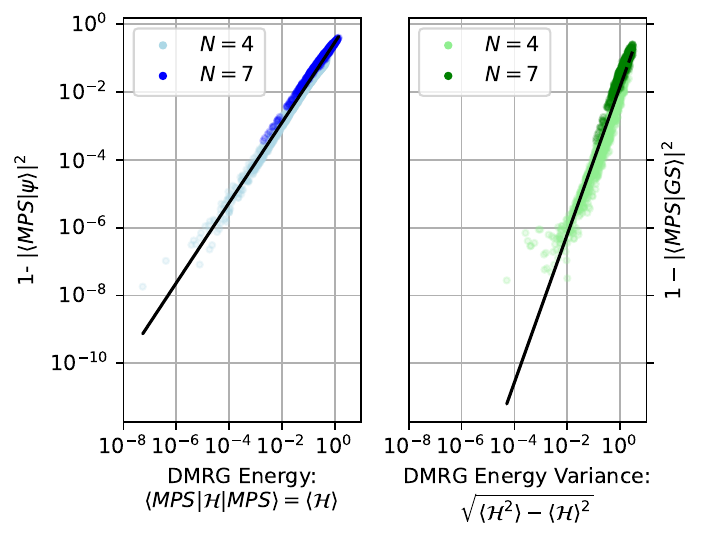}}
\caption{By fitting a line to each of the datasets in all of the log-log plots we obtain the above power-law relations.
}
\end{figure*}
\newpage
\subsection{\label{appendix:secAlgebraicMPOs}Algebraic MPO construction}

The goal is to construct the MPO representation of eq. \ref{eq:nonGBSHamiltonian}. Without loss of generality we can set $\pmb{\mu}=\pmb{0}=\pmb{\nu}$ because it is a simple matter to introduce them later, through a redefinition of the quadrature operators (excluding those that stem from $\pmb{\hat{\xi}}$); $\hat{X} \rightarrow \hat{X}-\mu$ and $\hat{P} \rightarrow \hat{P} - \nu$.
While it is not immediately clear how to write up the MPO for the full Hamiltonian, we begin by writing out the Hamiltonian as: 
\begin{align}
    \mathcal{H} = {\color{red} \mathcal{H}_\text{Gaussian}}
    &+ {\color{purple} \kappa\sum_i\beta_{ii}\hat{P}_i\prod_{n\neq i} \hat{X}_n
    + \kappa^2\sum_i b_{ii}\prod_{n\neq i} \hat{X}_n^2} \label{appendix:eqNonGaussianHamiltonian}\\
    &+ {\color{blue} \kappa\sum_{i\neq j} \prod_{n\neq i,j}\hat{X}_n(\gamma_{ij}\hat{X}_i^2 + b_{ij}(\hat{X}_i\hat{P}_i + \hat{P}_i\hat{X}_i)) + \kappa \text{Tr}\gamma\prod_n \hat{X}_n}
    + {\color{magenta} \kappa^2\sum_{i>j}\beta_{ij}\hat{X}_i\hat{X}_j\prod_{n\neq i,j}\hat{X}_n^2}\nonumber
\end{align}
The color coding is for more easily matching MPOs to Hamiltonian terms, also recall that this $\kappa$ is due to eq. \ref{eq:GlobalCZgate}, not eq. \ref{eq:pLBOUnitary}.
Here, $\mathcal{H}_\text{Gaussian}$ is eq. \ref{eq:GBSHamiltonian} and 
\begin{equation}
    \alpha_{ij} := \frac12(V^{-1})_{ij}
    \quad\quad 
    \beta_{ij} := \frac12(V^{-1})_{(N+i)(N+j)} =: 2b_{ij}
    \quad\quad
    \gamma_{ij} := \frac12(V^{-1})_{(i)(N+j)}
\end{equation}
The approach will be to write each term as an MPO and write the MPO of the full Hamiltonian as a block-diagonal MPO. This way, it is easier to derive the form of the MPO while we can still utilize the sparseness of the full MPO in the computations during optimization. 
We will for pedagogical reasons assume an odd number of modes, but it is straightforward to infer the MPOs for the case of an even number of modes. The index for the middle-most mode is denoted $N_m = (N+1)/2$.
All the MPOs will be of the form:
\begin{equation}
    \text{MPO} : \left(\prod_{s=1}^{N_m-1} L_s\right) M_{N_m}\prod_{s=N_m+1}^{N}R_s
\end{equation}
where the $L_s$', $M_{N_m}$ and $R_s$' are matrices containing operators that only act in the Hilbert space corresponding to the sub-index $s$. The idea behind every construction is to use the $L_s$' and $R_s$' to collect the relevant operator strings and then collapse the vector-matrix-vector product with the right coefficients in $M_{N_m}$, in order to yield all the terms in the Hamiltonian.
This will become clearer by considering the explicit construction of the matrices/MPO-cores in the following.

\subsubsection{{\color{red} $\mathcal{H}_\text{Gaussian}$ MPO}}
Firstly, we define the single site operator:
\begin{equation}
    \hat{S}_i = \frac{\alpha_{ii}}{2}\hat{X}^2_i + \frac{\beta_{ii}}{2}\hat{P}^2_i+\frac{\gamma_{ii}}{2}\left(\hat{X}_i\hat{P}_i+\hat{P}_i\hat{X}_i\right) - \frac{1}{2}
\end{equation}
such that the cores can be written as:
\begin{align}
    L_1 &=
    \begin{pmatrix}
        \hat{X}_1 & \hat{P}_1 & \mathbbm{1} & \hat{S}_1
    \end{pmatrix}\\
    R_N^T &= 
    \begin{pmatrix}
        \hat{X}_N & \hat{P}_N & \mathbbm{1} & \hat{S}_N
    \end{pmatrix}\\
    R_{s\neq N} &= 
    \begin{pmatrix}
        0 & 0 & & 0 & \hat{X}_s & 0 \\
        0 & 0 & \hdots & 0 & \hat{P}_s & 0 \\
        \mathbbm{1} & 0 & & 0 & 0 & 0\\
        0 & \mathbbm{1} & & 0 & 0 & 0\\
        \vdots &   & \ddots &  & \vdots & \\
        0 & 0 &  & \mathbbm{1} & 0 & 0 \\
        0 & 0 & \hdots & 0 & \mathbbm{1} & 0\\
        {\begin{aligned}&\hat{X}_s\alpha_{s(s+1)} \\&+ \hat{P}_s\gamma_{(s+1)s}\end{aligned}} & {\begin{aligned}&\hat{X}_s\gamma_{s(s+1)} \\&+ \hat{P}_s\beta_{s(s+1)}\end{aligned}} &  & {\begin{aligned}&\hat{X}_s\gamma_{sN} \\&+ \hat{P}_s\beta_{sN}\end{aligned}} & \hat{S}_s & \mathbbm{1}
    \end{pmatrix}\\
    M_{N_m} &=
    \begin{pmatrix}
        \alpha_{(N_m-1)(N_m+1)} & \gamma_{(N_m-1)(N_m+1)} &  & \alpha_{(N_m-1)N} & \gamma_{(N_m-1)N} & \alpha_{(N_m-1)N_m}\hat{X}_{N_m} + \gamma_{(N_m-1)N_m}\hat{P}_{N_m} & 0\\
        \gamma_{(N_m+1)(N_m-1)} & \beta_{(N_m-1)(N_m+1)} & \hdots & \gamma_{N(N_m-1)} & \beta_{(N_m-1)N} & \gamma_{N_m(N_m-1)}\hat{X}_{N_m} + \beta_{(N_m-1)N_m}\hat{P}_{N_m} & 0 \\
         & \vdots & \ddots &  &  & \vdots  & \\
        \alpha_{1(N_m+1)} & \gamma_{1(N_m+1)} & &\alpha_{1N} & \gamma_{1N} & \alpha_{1N_m}\hat{X}_{N_m} + \gamma_{1N_m}\hat{P}_{N_m} & 0 \\
        \gamma_{(N_m+1)1} & \beta_{1(N_m+1)} & & \gamma_{N1} & \beta_{1N} & \gamma_{N_m1}\hat{X}_{N_m} + \beta_{1N_m}\hat{P}_{N_m} & 0 \\
        {\begin{aligned}&\hat{X}_{N_m}\alpha_{N_m(N_m+1)} \\&+ \hat{P}_{N_m}\gamma_{(N_m+1)N_m}\end{aligned}} & {\begin{aligned}&\hat{X}_{N_m}\gamma_{N_m(N_m+1)} \\&+ \hat{P}_{N_m}\beta_{N_m(N_m+1)}\end{aligned}} & \hdots &
        {\begin{aligned}&\hat{X}_{N_m}\alpha_{N_mN} \\&+ \hat{P}_{N_m}\gamma_{NN_m}\end{aligned}} & {\begin{aligned}&\hat{X}_{N_m}\gamma_{N_mN} \\&+ \hat{P}_{N_m}\beta_{N_mN}\end{aligned}} &
        \hat{S}_{N_m} & \mathbbm{1}\\
        0 & 0 & & 0 & 0 & \mathbbm{1} & 0
    \end{pmatrix}
\end{align}
The $L_s$ are similar to the $R_s^T$ but,
\begin{align*}
\text{with the row:}\quad\quad\quad &\begin{pmatrix}
    {\begin{aligned}&\hat{X}_s\alpha_{s(s+1)} \\&+ \hat{P}_s\gamma_{(s+1)s}\end{aligned}} & {\begin{aligned}&\hat{X}_s\gamma_{s(s+1)} \\&+ \hat{P}_s\beta_{s(s+1)}\end{aligned}} & \hdots & {\begin{aligned}&\hat{X}_s\alpha_{s(s+1)} \\&+ \hat{P}_s\gamma_{(s+1)s}\end{aligned}}& {\begin{aligned}&\hat{X}_s\gamma_{sN} \\&+ \hat{P}_s\beta_{sN}\end{aligned}} & \hat{S}_s & \mathbbm{1}
\end{pmatrix}\\
\text{replaced by:}\quad\quad\quad &\begin{pmatrix}
    {\begin{aligned}&\hat{X}_s\alpha_{s(s-1)} \\&+ \hat{P}_s\gamma_{(s-1)s}\end{aligned}} & {\begin{aligned}&\hat{X}_s\gamma_{s(s-1)} \\&+ \hat{P}_s\beta_{s(s-1)}\end{aligned}} & \hdots & {\begin{aligned}&\hat{X}_s\alpha_{s1} \\&+ \hat{P}_s\gamma_{1s}\end{aligned}}& {\begin{aligned}&\hat{X}_s\gamma_{s1} \\&+ \hat{P}_s\beta_{s1}\end{aligned}} & \hat{S}_s & \mathbbm{1}
\end{pmatrix}
\end{align*}
\noindent\fbox{\begin{minipage}{\textwidth}
As an example, consider $N=5$:
\begin{equation}
    \includegraphics[width=1\linewidth,trim={0cm 5cm 0cm 8cm},clip]{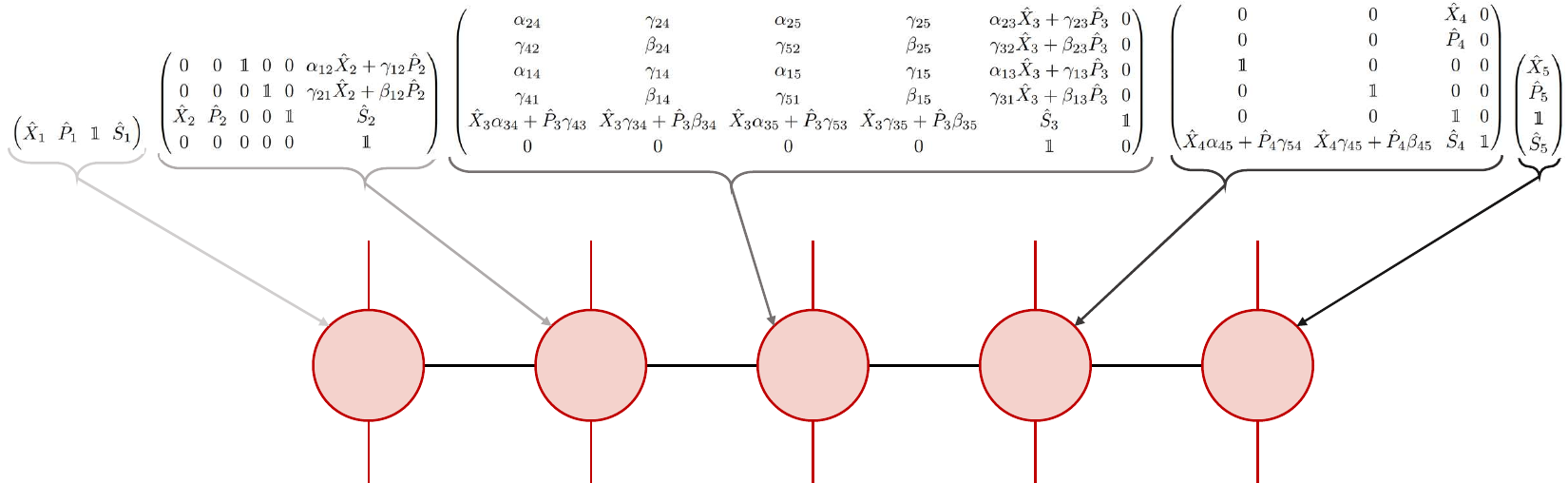}
\end{equation}
Doing the matrix multiplication gives all the terms of the Hamiltonian used for Gaussian Boson Sampling:
\begin{equation}
    {\color{red} \mathcal{H}_\text{Gaussian}} = \frac{1}{2}
    \begin{pmatrix}
        \pmb{\hat{X}} & \pmb{\hat{P}}
    \end{pmatrix}
    \left(
    \begin{array}{c|c} 
      \alpha & \gamma \\ 
      \hline 
      \gamma^T & \beta
    \end{array} 
    \right)
    \begin{pmatrix}
        \pmb{\hat{X}}\\
        \pmb{\hat{P}}
    \end{pmatrix} - \frac{N}{2} 
\end{equation}
\end{minipage}}

\subsubsection{{\color{purple} Second and third terms as MPOs} }
The MPOs of the second and third terms are more easily written up as:
\begin{align}
\sum_i\beta_{ii}\hat{P}_i\prod_{n\neq i} \hat{X}_n &= 
    \begin{pmatrix}
        \beta_{11}\hat{P}_1 & \hat{X}_1
    \end{pmatrix}
    \left(
    \prod_{n=2}^{N-1}
    \begin{pmatrix}
        \hat{X}_n & 0 \\
        \beta_{nn}\hat{P}_n & \hat{X}_n
    \end{pmatrix}
    \right)
    \begin{pmatrix}
        \hat{X}_N \\
        \beta_{NN}\hat{P}_N
    \end{pmatrix}\\
\sum_i b_{ii}\prod_{n\neq i} \hat{X}_n^2 &=
    \begin{pmatrix}
        b_{11} & \hat{X}^2_1
    \end{pmatrix}
    \left(
    \prod_{n=2}^{N-1}
    \begin{pmatrix}
        \hat{X}^2_n & 0 \\
        b_{nn} & \hat{X}^2_n
    \end{pmatrix}
    \right)
    \begin{pmatrix}
        \hat{X}^2_N \\
        b_{NN}
    \end{pmatrix}
\end{align}

\subsubsection{{\color{blue} The MPO of the fourth and fifth term}}
Now, we consider the explicit construction of the following MPO:
\begin{equation}
    \sum_{i\neq j} \prod_{n\neq i,j}\hat{X}_n(\gamma_{ij}\hat{X}_i^2 + b_{ij}(\hat{X}_i\hat{P}_i + \hat{P}_i\hat{X}_i))
    + \text{Tr}\gamma \prod_n \hat{X}_n= \left(\prod_{s=1}^{N_m-1} L_s\right) M_{N_m}\prod_{s=N_m+1}^{N}R_s
\end{equation}
To this end we define: $\Tilde{P}_n = \hat{X}_n\hat{P}_n + \hat{P}_n\hat{X}_n$ 
such that the cores can be written as:
\begin{align}
    &L_1 =
    \begin{pmatrix}
        \mathbbm{1} & \hat{X}_1 & \hat{X}_1^2 & \Tilde{P}_1
    \end{pmatrix}\\
    &R_N^T =
    \begin{pmatrix}
        \mathbbm{1} & \hat{X}_N & \hat{X}_N^2 & \Tilde{P}_N
    \end{pmatrix}\\
    &R_s = 
    \begin{pmatrix}
        \hat{X}_s & 0 & \hdots \\
        0 & \ddots &  \\
        \vdots &  & \hat{X}_s & 0 & 0 & 0\\
         & & 0 & \mathbbm{1} & 0 & 0\\
         & & 0 & \hat{X}_s & 0 & 0\\
         & & 0 & \hat{X}_s^2 & 0 & 0\\
         & & 0 & \Tilde{P}_s & 0 & 0\\
         & & 0 & 0 & \hat{X}_s& 0\\
         & & 0& 0& 0& \hat{X}_s \\
         & & & & & & \ddots\\
         & & & & & & 0 & \hat{X}_s & 0\\
         \gamma_{sN}\hat{X}_s^2+b_{sN}\Tilde{P}_{s} & \hdots& \gamma_{s(s+1)}\hat{X}_s^2 +b_{s(s+1)}\Tilde{P}_{s}& 0 &\gamma_{(s+1)s} &b_{(s+1)s} &\hdots & b_{Ns}& \hat{X}_s
    \end{pmatrix}\\
&\resizebox{1.1\hsize}{!}{$
    M_{N_m} = 
    \begin{pmatrix}
        0 & \hdots & 0 & \gamma_{N_m1}\hat{X}_{N_m}^2 + b_{N_m1}\Tilde{P}_{N_m} & \gamma_{(N_m+1)1}\hat{X}_{N_m} & b_{(N_m+1)1}\hat{X}_{N_m} & \hdots & \gamma_{N1}\hat{X}_{N_m} & b_{N1}\hat{X}_{N_m} & 0 \\
        \vdots & \ddots &  & \vdots & & & \ddots &  \\
        0 &  & 0 & \gamma_{N_m(N_m-1)}\hat{X}_{N_m}^2 + b_{N_m(N_m-1)}\Tilde{P}_{N_m} & \gamma_{(N_m+1)(N_m-1)}\hat{X}_{N_m} & b_{(N_m+1)(N_m-1)}\hat{X}_{N_m} & \hdots & \gamma_{N(N_m-1)}\hat{X}_{N_m} & b_{N(N_m-1)}\hat{X}_{N_m} & 0 \\
        \gamma_{N_mN}\hat{X}_{N_m}^2 + b_{N_mN}\Tilde{P}_{N_m} & \hdots & 
        \gamma_{N_m(N_m+1)}\hat{X}_{N_m}^2 + b_{N_m(N_m+1)}\Tilde{P}_{N_m}
        & \text{Tr}\gamma \hat{X}_{N_m} & \gamma_{(N_m+1)N_m} & b_{(N_m+1)N_m} & \hdots & \gamma_{NN_m} & b_{NN_m} & \hat{X}_{N_m} \\
        \gamma_{(N_m-1)N}\hat{X}_{N_m} & & \gamma_{(N_m-1)(N_m+1)}\hat{X}_{N_m} & \gamma_{(N_m-1)N_m} & 0 & 0 \\
        b_{(N_m-1)N}\hat{X}_{N_m} & & b_{(N_m-1)(N_m+1)}\hat{X}_{N_m} & b_{(N_m-1)N_m} & 0 & \ddots \\
        \vdots & \ddots & \vdots \\
        \gamma_{1N}\hat{X}_{N_m} & & \gamma_{1(N_m+1)}\hat{X}_{N_m} & \gamma_{1N_m} \\
        b_{1N}\hat{X}_{N_m} & & b_{1(N_m+1)}\hat{X}_{N_m} & b_{1N_m} & & &  & &\ddots &0\\
        0 & \hdots & 0 & \hat{X}_{N_m} & & & && 0& 0
    \end{pmatrix}$}
\end{align}
The $L_s$ are similar to the $R_s^T$ but,
\begin{align*}
\text{with the row:}\quad\quad\quad &\begin{pmatrix}
    \gamma_{sN}\hat{X}_s^2+b_{sN}\Tilde{P}_{s}& \hdots& \gamma_{s(s+1)}\hat{X}_s^2+b_{s(s+1)}\Tilde{P}_{s}& 0 &\gamma_{(s+1)s} &b_{(s+1)s} &\hdots & b_{Ns}& \hat{X}_s
\end{pmatrix}\\
\text{replaced by:}\quad\quad\quad &\begin{pmatrix}
    \gamma_{s1}\hat{X}_s^2++b_{s1}\Tilde{P}_{s}& \hdots& \gamma_{s(s-1)}\hat{X}_s^2++b_{s(s-1)}\Tilde{P}_{s}& 0 &\gamma_{(s-1)s} &b_{(s-1)s} &\hdots & b_{1s}& \hat{X}_s
\end{pmatrix}
\end{align*}

\subsubsection{{\color{magenta} The MPO for the last term}}
Again, we write the MPO as:
\begin{equation}
    \sum_{i>j}\beta_{ij}\hat{X}_i\hat{X}_j\prod_{n\neq i,j}\hat{X}_n^2 = \left(\prod_{s=1}^{N_m-1} L_s\right) M_{N_m}\prod_{s=N_m+1}^{N}R_s
\end{equation}
this time with: 
\begin{align}
    &L_1 =
    \begin{pmatrix}
        \hat{X}_1^2 & \hat{X}_1
    \end{pmatrix}\\
    &R_N^T =
    \begin{pmatrix}
        \hat{X}_N^2 & \hat{X}_N
    \end{pmatrix}\\
    &R_s = 
    \begin{pmatrix}
        \hat{X}_s^2 & 0 \\
        \hat{X}_s & 0 \\
        0 & \hat{X}_s^2  \\
        & & \ddots \\
        & & &\hat{X}_s^2 \\
        0 & \beta_{(s+1)s}\hat{X}_s & \hdots & \beta_{Ns}\hat{X}_s & \hat{X}_s^2
    \end{pmatrix}\\
    &M_{N_m} = 
    \begin{pmatrix}
        0 & \beta_{(N_m+1)N_m}\hat{X}_{N_m} & \hdots & \beta_{NN_m}\hat{X}_{N_m} & \hat{X}_{N_m}^2 \\
        \beta_{N_m(N_m-1)}\hat{X}_{N_m} & \beta_{(N_m+1)(N_m-1)}\hat{X}_{N_m}^2 & \hdots & \beta_{N(N_m-1)}\hat{X}_{N_m}^2 & 0 \\
        \vdots  & \vdots & \ddots &\vdots & \vdots \\
        \beta_{N_m1}\hat{X}_{N_m} & \beta_{(N_m+1)1}\hat{X}_{N_m}^2 & \hdots & \beta_{N1}\hat{X}_{N_m}^2 & 0 \\
        \hat{X}_{N_m}^2 & 0&\hdots & 0& 0
    \end{pmatrix}
\end{align}
The $L_s$ are similar to the $R_s^T$ but,
\begin{align*}
\text{with the row:}\quad\quad\quad &\begin{pmatrix}
    0 & \beta_{(s+1)s}\hat{X}_s & \hdots & \beta_{Ns}\hat{X}_s & \hat{X}_s^2
\end{pmatrix}\\
\text{replaced by:}\quad\quad\quad &\begin{pmatrix}
    0 & \beta_{s(s-1)}\hat{X}_s & \hdots & \beta_{s1}\hat{X}_s & \hat{X}_s^2
\end{pmatrix}
\end{align*}
\end{widetext}

\bibliography{biblio}

\end{document}